\documentclass[preprint2]{aastex62-mod}

\graphicspath{{./}{figures/}}

\accepted{Dec 19, 2018 by The Astrophysical Journal, Post-print version.\\ PuPublished Feb 8, 2019, doi: 10.3847/1538-4357/aafd30}

\shorttitle{Magnetic turbulence spectra in the IHS and LISM}
\shortauthors{Fraternale et al.}
\usepackage{amsmath}
\usepackage{amsbsy}
\usepackage{amsfonts}
\usepackage{amssymb}
\usepackage{bm}

\usepackage{placeins}
\usepackage{xcolor}
\definecolor{orangered}{HTML}{FF7110}
\definecolor{green}{HTML}{00CC00}

\turnoffediting

\begin{document}

\title{Magnetic turbulence spectra and intermittency in the heliosheath and in the local interstellar medium}

\correspondingauthor{Daniela Tordella}
\email{daniela.tordella@polito.it}

\author[0000-0002-4700-2762]{Federico Fraternale}
\affil{Dipartimento di Scienza Applicata e Tecnologia\\Politecnico di Torino\\
10129, Torino, Italy}

\author{Nikolai V. Pogorelov}
\affiliation{Department of Space Science\\ University of Alabama in Huntsville\\ Huntsville, AL 35805, USA}
\altaffiliation{Center for Space Plasma and Aeronomic Research\\ University of Alabama in Huntsville\\ Huntsville, AL 35805, USA}

\author{John D. Richardson}
\affiliation{Kavli Institute for Astrophysics and Space Research\\ Massachusetts Institute of Technology\\ Cambridge (MA), 02139, USA}

\author{Daniela Tordella}
\affil{Dipartimento di Scienza Applicata e Tecnologia\\Politecnico di Torino\\
10129, Torino, Italy}



\begin{abstract}
The understanding of inertial-scale dynamics in the heliosheath is not yet thorough. Magnetic field fluctuations across the inner heliosheath and the local interstellar medium are here considered to provide accurate and highly resolved statistics over different plasma conditions between 88 and 136 AU. 
By using the unique \textit{in situ} 48-s  measurements from the \textit{Voyager Interstellar Mission}, we investigate different fluctuation regimes at the magnetohydrodynamic (MHD) scales, down to the MHD-to-kinetic transition. We focus on a range of scales exceeding five frequency decades ($5\times10^{-8}<f<10^{-2}$ Hz), which is unprecedented in literature analysis. A set of magnetic field data for eight intervals in the inner heliosheath, in both unipolar and sector regions, and four intervals in the local interstellar medium is being used for the analysis. Results are set forth in terms of the power spectral density, spectral compressibility, structure functions and intermittency of magnetic field increments. In the heliosheath, we identify the \textit{energy-injection} regime displaying a $\sim1/f$ energy decay, and the \textit{inertial-cascade} regime. Here, the power spectrum is anisotropic and dominated by compressive modes, with intermittency that can reach kurtosis values up to ten. In the interstellar medium the structure of turbulence is anisotropic as well, with transverse fluctuations clearly prevailing after May 2015. Here, we show that intermittent features occur only at scales smaller than $10^{-6}$ Hz.

\end{abstract}



\section{Introduction} \label{sec:intro}

\emph{Voyagers}~1 and~2 (\textit{V1} and \textit{V2}) crossed the heliospheric termination shock (TS) in December~2004  and in August~2007,
respectively \citep{stone2005,stone2008}. It is widely accepted that on 25 August 2012, \textit{V1} crossed the heliopause (HP) and is now moving through the local interstellar medium (LISM) \citep{burlaga2013,burlaga2014c,gurnett2013,stone2013,webber2013}.



The magnetic field behavior beyond the HP is determined by
the interstellar magnetic field (ISMF) draping over the HP as a tangential discontinuity that separates the solar wind (SW) from the LISM. The simulations of \citet{pogorelov2009,pogorelov2017b,borovikov2014} predicted the magnetic field behavior very similar to observations. The recent simulation of \citet{kim2017} shows that the ISMF ``undraping''
is consistent with the time-dependent processes occurring in the SW. It also reproduces major shocks propagating through the LISM that cause the plasma wave events observed by PWS  between Nov 2012 to Jul 2016 \citep{gurnett2013,gurnett2015,pogorelov2017b}. Interestingly, it also predicted a shock passing through \textit{V1} in Aug 2017,  when plasma waves were observed by PWS.
\textit{V2} remains in the inner heliosheath (IHS) measuring a velocity profile very different from that of \textit{V1} at the same distance from the Sun and gradually approaches the HP. Conventionally, the IHS is the SW region between the TS and the HP. The heliopause has ``structure'' clearly observed by \textit{V1}. It was crossed in about 1 month, which gives width of about 0.3~AU. There are strong indications that this ``structure'' is due to the HP instability \citep{borovikov2014} or magnetic reconnection \citep{schwadron2013,pogorelov2017b}, or both.

Theoretical and numerical studies of the SW--LISM interaction have a long history reviewed, e.g., in \cite{zank1999,zank2015,izmodev2009,izmodev2015,opher2016}, and \cite{pogorelov2017a}.
The TS is formed due to deceleration of the supersonic wind when it interacts with the HP and LISM counter-pressure.
Modern models of the SW--LISM interaction take into account the effects of charge-exchange between ions and neutral atoms,coupling of the heliospheric magnetic field (HMF) and ISMF, and treat non-thermal (pickup) ions (PUIs) as a separate component \citep[see][and references therein]{pogorelov2016,pogorelov2017b}.

As far as the model uncertainties are concerned, remarkably many observations of the SW/LISM bulk flow and average magnetic field have been reproduced by simulations. E.g., the deflection of the LISM neutral H atoms is on average in the $BV$-plane, which is defined by the LISM velocity and ISMF vectors, $\mathbf{V}_\infty$  and $\mathbf{B}_\infty$, in the unperturbed LISM \citep{izmodev2005,pogorelov2008,pogorelov2009b,katushkina2015}. Kinetic energetic neutral atom (ENA) flux simulations of \citet{heerikhuisen2010,heerikhuisen2014,heerikhuisen2011,zirnstein2015,zirnstein2015b,zirnstein2016,zirnstein2017,zirnstein2018} reproduced the \textit{IBEX} ENA ribbon using the $BV$-plane consistent with the hydrogen deflection plane.  A number of simulations reproduce \textit{Voyager} measurements. \citet{pogorelov2017b} demonstrated that the distribution of density in the heliospheric boundary layer (a region of decreased plasma density on the LISM side of the HP) is in agreement with PWS data. These models also reproduce the H density at the TS derived from PUI measurements \citep{bzowski2009} and observed anisotropy in the 1--10 TeV galactic cosmic ray (GCR) flux \citep{schwadron2014,zhang2014,zhang2016}.

There are certain details in the observations and simulations that clearly demonstrate that turbulence bears an imprint of physical processes occurring in the IHS and LISM. These are related to the turbulent character of the SW both in front of the TS and in the IHS, the presence of the heliospheric current sheet (HCS) that separates the sectors of opposite HMF polarity, the variability of the boundary between the sector and unipolar HMF, and the possibility that instabilities and magnetic reconnection destroy the HCS, thus  resulting in the HMF decrease. In fact, simulations imply that turbulence may be affecting the interaction pattern by facilitating magnetic reconnection and instabilities. Moreover, the LISM turbulence may be affected by shocks propagating through it.

\textcolor{black}{This paper is an attempt to address these issues by performing a turbulence analysis of \textit{Voyager} data. In particular, this study provides a spectral characterization of the various regimes of magnetic field fluctuations from the \textit{energy-injection} range, through the \textit{inertial cascade},  down to scales where kinetic effects start to affect the dynamics ($\approx 10^4$ km). The analysis cannot be further extended to smaller scales due to resolution- and accuracy-related issues of \textit{Voyager} data. However, it is known that the inertial-cascade regime of turbulence keeps track of physical processes taking place at smaller scales, which makes its analysis significantly intriguing.}

\textbf{HMF sectors in the IHS.} 
A characteristic property of the SW flow is the existence of a HCS that separates magnetic field lines of opposite polarity, which originate at the solar surface. The HCS propagates with the solar wind kinematically, provided that it has no back reaction on the flow.
Theoretical and numerical, kinetic and multi-fluid analyses of magnetic reconnection across the HCS, have been discussed by \citet{drake2010,drake2017} and \citet{pogorelov2013a,pogorelov2017b}. Magnetic reconnection may reveal itself as a tearing mode (or plasmoid) instability and may take place especially close to the HP, where the sector width decreases to negligible values. 

Resolution of the sectors of alternating magnetic field polarity in the IHS is impossible. It is known that \textit{Voyager~1} had been observing negative radial velocity component for two years before it crossed the HP \citep{decker2012,pogorelov2009,pogorelov2012}. Because of the piling up effect, the sector width should be negligible near the HP.
Moreover, the sector width decreases to zero at the HCS tips, which makes attempts to resolve the traditional HCS structure very challenging.
Besides, current sheets can be created not only due to the tilt of the Sun's magnetic axis, but also due to stream interactions abundant in the heliosphere. The transition to chaotic behavior of the HMF 
occurs when the sector width becomes less than the numerical resolution \citep{pogorelov2017b}. In nature, this is possible both due to pre-existing turbulence and magnetic reconnection across the current sheets, which also creates turbulence.


The approach followed by \citet{borovikov2011} to track the HCS surface  based on the assumption of HMF being unipolar is neither practical, nor acceptable. Numerical simulations allow us to determine what happens to $\textbf{B}$ if the HMF is assumed to be unipolar: it is clear that the calculated magnetic field strength in this case is substantially overestimated as compared with \textit{V1} observations in the IHS \citep[see the discussions, e.g., in][]{pogorelov2015,pogorelov2017b}. Thus, the possibility of HMF depressions in the IHS covered by a sectored HMF should not be disregarded. It can be identified by the increased turbulence level in relevant regions. \textit{Voyager} spacecraft provide us with appropriate measurements to answer these questions.

\citet{richardson2016} have investigated the effect of the magnetic axis tilt on the number of HCS crossings and compared the observed and expected numbers. It has been shown that the number of HCS crossings substantially decreased two years after \textit{V1} and \textit{V2} entered the IHS. However, \textit{V2} might have entered the unipolar region at that time. It was concluded that there are indications of magnetic field decrease possibly due to magnetic reconnection across the HCS. In addition, as shown by \citet{drake2017}, \textit{V2} data reveal that fluctuations in the density and magnetic field strength are anticorrelated in the sector regions, as is expected from their magnetic reconnection modeling, but not in the unipolar regions. A possible dissipation of the HMF in such regions may also be an explanation of a sharp reduction in the number of sectors, as seen from the \emph{V1} data.

\textbf{Turbulence in the heliosheath and LISM.}
An extensive data analysis related to the SW turbulence behavior in the SW ahead of the TS and in the IHS along the \textit{Voyager} trajectories was performed by
\citet{burlaga1994, burlaga2003a,burlaga2003b,burlaga2003c,burlaga2006,burlaga2007,burlaga2009,burlaga2009b,burlaga2015,burlaga2017,burlaga2018}.
There is no single physical mechanism responsible for all observed turbulence manifestations. Large-amplitude fluctuations in the magnetic field strength $B$ are observed at small scales with very complex profiles. These fluctuations were described as ``turbulence'' \citep{burlaga2006,fisk2008},
although their nature and origin are not yet understood.
The turbulence includes ``kinetic scale'' features (with sizes of the order of 10--100 gyroradii) and microscale features ($>100$ proton gyroradii).
Usually, the observed turbulence consists of both coherent and random structures as seen in time profiles of the magnetic field strength on scales from 48 s to several hours.
As shown by \citet{burlaga2009b}, the large-scale (1 day) fluctuations measured at \textit{V1} in the unipolar  and sector regions differ in some aspects: they have a log-normal distribution in the post-TS region, but Gaussian in the unipolar region. Instabilities and magnetic reconnection enhance turbulence, i.e, magnetic field dissipation in the sector region should naturally affect magnetic field and plasma density fluctuations inside the IHS.

Magnetic field fluctuations have been observed also in the LISM. However, those fluctuations are smaller than in the IHS \citep{burlaga2018} and their nature is not yet understood. In this paper, we  investigate the LISM turbulence in different regions separated by shocks causing plasma wave emission observed by \textit{V1}. Our methodology allows us to investigate these fluctuations in more detail than earlier. 
In fact, as discussed in \S\ref{sec:data}, analysis of power spectra in the SW turbulence necessary to address these issues is a challenging task because of the sparsity of the 48 s data. After the TS, about 70\% of magnetic field 48-s data are missing.  Thus, sophisticated spectral estimation techniques become mandatory for obtaining reliable and physically meaningful results. 

In the present study, Section \ref{sec:data} describes the data sets used for the analysis. Section \ref{sec:IHS} contains results of the analysis of the inner heliosheath, and it is split in two parts: \S \ref{sec:V2_IHS} for \textit{Voyager 2 } and \S \ref{sec:V1_IHS} for \textit{Voyager 1}. LISM turbulence is discussed in Section \ref{sec:LISM}, and final remarks follow in Section \ref{sec:conclusions}. In  Appendix \ref{sec:App_A}, we report information of the methods used for spectral estimation. Finally,  Appendix \ref{sec:App_B}, contains information on variance anisotropy. 

\section{Voyager data in the IHS and LISM and selected periods.}\label{sec:data}
This study considers different intervals in the inner heliosheath and in the local interstellar medium. In particular,  we used magnetic field data at the highest resolution publicly available, the 48-s averaged data measured \textit{in situ} by the Voyager Interstellar Mission (\url{https://voyager.jpl.nasa.gov/mission/interstellar-mission/}). In the heliosheath and beyond, the \textit{Voyager} LMF magnetometers (MAG experiment, see  \cite{behannon1977}) sample magnetic field at a rate of 2.08 samples per second. The rate of the telemetry is 0.0208 Hz, and 48 s averages are periodically published in the NASA Space Physics Data Facility (\url{https://spdf.gsfc.nasa.gov/}) and can also be accessed via the COHO web site (\url{https://cohoweb.gsfc.nasa.gov/coho/}). Data are currently available through day-of-year (DOY) 271 of 2017 for V1, and through 2015 DOY 356 for V2, in RTN reference frame.
The Heliographic RTN coordinate system is centered at the spacecraft. The R axis points radially outward from the Sun, the T axis is parallel to the solar equatorial plane and points in the direction of the Sun's rotation, while the N axis completes the orthonormal triad.  
 In the most sensitive LMF range, the level of noise is 0.006 nT. However, 1-sigma systematic errors due to the data calibration process and other sources of noise (sensors, electronics, telemetry system, ground tracking stations) are estimated around $\pm$0.02 and $\pm$0.03 nT at V1 and V2, respectively. Besides, the variability of errors makes such uncertainties rise up to $\pm0.1$ nT in specific periods or for specific field components  \citep{berdichevsky2009}. In addition to the noise, limited telemetry  coverage (the Camberra antennas of the CDSCC can only view V1 and V2 twelve hours per day) leads to data gaps of 8-16 hours per day. This point constitutes the major challenge for a spectral analysis, such as that presented for the first time in this study. The numerical techniques we used are synthetically described in Appendix \ref{sec:App_A}, and previously used in \cite{fraternale2016,gallana2016,fraternale2017phd}.

\begin{figure}
\includegraphics[width=0.48\textwidth]{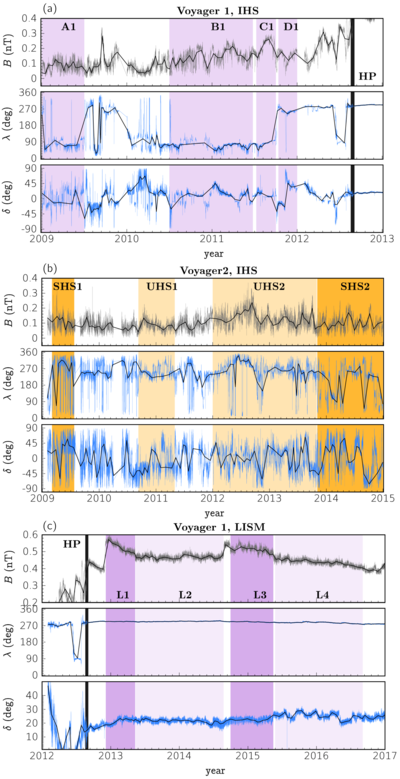}
    \caption{Data sets analyzed in this study. Top panels (a): V1, IHS. Middle panels (b): V2, IHS. Bottom panels (c): V1, LISM. Each panel contains, from top to bottom, the magnetic field magnitude, $B=|\mathbf{B}|$; the azimuthal angle, $\lambda=\tan^{-1}({B_\mathrm{T}}/ {B_\mathrm{R}})$; the elevation angle, $\delta=\sin^{-1}({B_\mathrm{N}}/ {B})$. Data points with $|B_\mathrm{R,T,N}|<0.03$ nT  have not been used in the computation of $\lambda$ and $\delta$. \label{fig:data} }
\end{figure}

In the IHS we consider four periods for V1 and for V2, respectively, after 2009 (see Figure \ref{fig:data}a). In particular, at V1 we selected the intervals: (\textbf{A1}) 2009, DOY 22 - 2009, DOY 180 (109.5$\pm0.77$ AU); (\textbf{B1}) 2010, DOY 180 - 2011, DOY 180 (115.65$\pm$1.79 AU); (\textbf{C1}) 2011, DOY 180 - 2011, DOY 276 (117.91$\pm$0.47 AU); (\textbf{D1}) 2011, DOY 276 - 2011, DOY 365 (118.81$\pm$0.44). During these periods, \textit{Voyager 1} sampled almost unipolar magnetic fields with northern ``toward'' polarity with respect to the Sun. Periods \textbf{C1} and \textbf{D1} are separated by a sector boundary crossing which occurred in 2011, DOY 276 and lasted about one day, when the  polarity became southern, ``away'', until 2012 DOY 209. During this period, interaction with the local interstellar plasma likely occurred. A  detailed description of sector  boundaries  in proximity of the heliopause from 2011.5 can be found in \cite{burlaga2014}.

As the Plasma Science (PLS) instrument is not operational at V1, the bulk wind speed sometimes can be recovered using a Compton-Getting analysis from Low Energy Charged Particle experiment (LECP) and Cosmic Ray Subsystem (CRS)  data \citep{krimigis2011}.  The heliosheath plasma has been provided by \citet{richardson2013,richardson2014}, and \citet{richardson2015}.  For our analysis, it is particularly important to highlight that V1 (traveling at $V_\mathrm{SC1}\approx 17$ km s$^{-1}$, 34.5$^\circ$ North) measured low radial velocity components since its crossing the heliospheric termination shock (TS). In particular, the radial velocity $V_\mathrm{R}$ decreased almost linearly from about 100 km s\textsuperscript{-1} (2006) to 0 km s\textsuperscript{-1} in 2010.5, while the tangential speed $V_\mathrm{R}$ oscillated around -40 km s\textsuperscript{-1}. Numerical simulations  \citep{pogorelov2013a} suggest that the absence of sector boundary crossings observed by V1 near the heliopause (HP), as well as the negative radial velocity observed during 2011 could be symptom of V1 being inside a \textit{magnetic barrier}.

The flow at \textit{Voyager 2} ($V_\mathrm{SC2}\approx15$  km s\textsuperscript{-1}, 30$^\circ$ South) was quite different. The bulk speed remained almost constant at about 150 km s\textsuperscript{-1} throughout the IHS. In contrast to V1 observations, high variability was found at V2 in the fluxes of energetic particles, as shown by \citet{decker2008}. Such variability has been related to the possibility for V2 to be alternatively inside the unipolar region (UHS) or inside the sector region (SHS) \citep{opher2011,hill2014}. The sector region is defined as the region swept by the heliospheric current sheet (HCS). Likely, V2 has been  very close to the boundary between these two regions, according to models based on kinematic propagation of the HCS's maximal extension, measured by the Wilcox Solar Observatory (\url{http://wso.stanford.edu/}). \cite{richardson2016} compared the results of two models with the actual number of sector boundary crossings observed at V2. It is believed that the spacecraft remained in the unipolar region from until 2013.83 when it entered the sector region, as discussed by \citet{burlaga2017}. 

Based on literature analysis, we selected four intervals for V2, see Figure \ref{fig:data}(b): (\textbf{SHS1}) 2009, DOY 62 - 2009, DOY 210 (89.0$\pm$0.65 AU); (\textbf{UHS1}) 2010 DOY 252 - 2011 DOY 210 (94.08$\pm$1.89 AU); (\textbf{UHS2}) 2012, DOY 1 -2013, DOY 300 (100.20$\pm$2.88 AU); (\textbf{SHS2}) 2013, DOY 300 - 2015, DOY 1 (104.95$\pm$1.87 AU). 
The thermal ions average plasma density ($n_i$) was about 0.001  cm\textsuperscript{-3} from 2009 to 2012, than increased to 0.002  cm\textsuperscript{-3} until 2015 (standard deviation is 0.0007 in the last period and 0.0003 in the earlier periods). The thermal plasma temperature was 63500 K in \textbf{SHS1}, 44700 K in \textbf{UHS1}, 56200 K in \textbf{UHS2},  and 51700 K in \textbf{SHS2} (standard deviation about 22000 K).

Ultimately, we considered four consecutive intervals of \textit{Voyager 1} data in the local interstellar medium (LISM), see Figure \ref{fig:data}c): (\textbf{L1}) 2012, DOY 340 - 2013, DOY 130 (123.3$\pm$0.77 AU); (\textbf{L2}) 2013 DOY 133 - 2014 DOY 236 (126.4 $\pm$2.29 AU); (\textbf{L3}) 2014 DOY 273 - 2015 DOY 135 (130.2$\pm$ 1.11 AU); (\textbf{L4}) 2015 DOY 145 - 2016 DOY 246 (133.7$\pm$ 2.28 AU). This partition was also identified by \citet{burlaga2016}: in that study, \textbf{L1} and \textbf{L3} were referred to as ``disturbed'' intervals while \textbf{L2} and \textbf{L4} as ``quiet''. The periods \textbf{L2} and \textbf{L4} have been also consedered by \citet{burlaga2015} and \citet{burlaga2018},  respectively. These regions are bounded by weak perpendicular shocks (or pressure waves) propagating through the LISM, as shown by \citet{burlaga2013c,gurnett2013,gurnett2015,burlaga2016}, and \citet{kim2017}.  The interstellar plasma is colder than the IHS ($T\approx10^4$ K). Electron plasma oscillations detected by V1's Plasma Wave Subsystem (PWS) yielded the density estimate of 0.08 cm\textsuperscript{-3}  \citep{gurnett2013}. These oscillations are driven by electron beams produced upstream of the shocks. Until 2016, five events have been detected (Oct-Nov 2012, Apr-May 2013, Feb-Nov 2014, Sep-Nov 2015). 

I this study, we removed outliers from the 48 s data sets. This was done by computing a backward and forward moving variance over a 48-hours window (3600 data points). For each magnetic field component, a data point was removed if  it was larger than 6 times the minimum between the backward and the forward moving standard deviation ($|B_j|>6 \min\{\sigma_{b_j},\sigma_{f_j}\}, j=1,\dots ,n$). Moreover, three calibration events (spacecraft rolls with 30 minutes periodicity) were removed from the intervals \textbf{A1} and \textbf{SHS1}. 
Magnetic field data have been rotated to mean field coordinates: a B-parallel component $B_\parallel$, and  two perpendicular components $B_{\perp1}$, $B_{\perp2}$ with respect to the average field $\mathbf{B_0}$, which better suits turbulence analysis \edit1{($B_{\perp1}$ is in the T-N plane and $B_{\perp2}$ completes the right-handed triplet)}. Since the  SW flow and the magnetic field directions in the IHS are nearly orthogonal, $B_\parallel\approx B_\mathrm{T}$, $B_{\perp1}\approx B_\mathrm{N}$ and $B_{\perp2}\approx B_\mathrm{R}$. Moreover, for each component, we removed linear trends.

In the following, bold-shaped letters indicate vector fields and non-bold letters are used for the magnitude of vector fields.

\section{Magnetic field fluctuations in the inner heliosheath}\label{sec:IHS}
This section shows the results of magnetic field fluctuation analysis in the IHS. We provide a spectral analysis for a frequency range wider than five decades, $f\in[10^{-8}, 10^{-2}]$ Hz. Before discussing results, few definitions and symbols are introduced. 

We computed the power spectral density ($\mathrm{PSD}$, or $P$), shown in Figures \ref{fig:V2_IHS_ene}, \ref{fig:V1_IHS_ene} for the IHS, and in left panels of Figure \ref{fig:V1_LISM_ene} for the LISM.  Due to the missing data issue, the PSD is estimated via three different numerical procedures (Appendix \ref{sec:App_A}). The comparative analysis of these techniques allows to recover the PSD with uncertainty of spectral indexes typically smaller 10\%.

We investigated the spectral compressibility \edit1{and variance anisotropy (Figures \ref{fig:spec_compressibility}, \ref{fig:LISM_compressibility}, \ref{fig:compress_anis_IHS_V2}, \ref{fig:compress_anis_IHS_V1}, \ref{fig:compress_anis_LISM_V1}). The anisotropy is expressed in terms of both $P[B_j]/E_m$ ($j=\{\parallel,\perp_1,\perp_2\}$) and $P[B_\perp]/P[B_\parallel]$, where $P[B_\perp]=P[B_{\perp1}]+P[B_{\perp2}]$. Due to the lack of accurate plasma data, as a proxy for spectral compressibility we use} the ratio between the PSD of the field magnitude and the trace: $C(f)=\mathrm{P}[B]/E_\mathrm{m}$, where $E_\mathrm{m}(f)=\mathrm{tr}(\mathrm{P}[\mathbf{B}])=\mathrm{P}[B_\parallel]+\mathrm{P}[B_{\perp1}]+\mathrm{P}[B_{\perp2}]$. 
This measure is an index of the alignment of the fluctuation vector with the average field.  \edit1{Fluctuations of the magnetic field magnitude can indeed be considered as a proxy for density fluctuations to a good degree of approximation. In fact, strong correlations between plasma density and $|\mathbf{B}|$ have been found in both rarefaction and compression regions between 1 and 11 AU  by \cite{roberts1987a}. Similar correlations had also been found previously by \cite{smith1983} and \cite{goldstein1983}. We also computed the fraction of parallel energy in the time domain, via two slightly different formulas:}  
\begin{gather}\label{eq:compressibility}
C_1=\left\langle\left(\frac{\mathbf{ B_0}\boldsymbol{\cdot}\boldsymbol{\delta}\mathbf{B}}{ B_0  \delta B}\right)^2\right\rangle,~
C_2=\frac{\langle(\mathbf{b}\boldsymbol{\cdot}\boldsymbol{\delta}\mathbf{B})^2\rangle}{\langle{\delta B^2}\rangle}.
\end{gather}
In the above expressions, $\delta \mathbf{B}=\mathbf{B}-\mathbf{B_0}$ is the three-dimensional fluctuation about the background field, $\delta B=(\delta B_\parallel^2+\delta B_{\perp1}^2+\delta B_{\perp2}^2)^{1/2}$ is its magnitude, 
$\mathbf{b}=\mathbf{ B_0}/B_0$ is the direction cosines vector. Angle brackets indicate the ensemble average over all the data points of each interval, and the dot indicates the scalar product. The powers of 2 used in Eq. \ref{eq:compressibility} allows us to interpret C1, C2 as the average percentage of fluctuating energy in the direction of $\mathbf{ B_0}$. To reduce the contribution of noise, the computation is performed with hourly-averaged data.  
Compressibility values are reported in Tables \ref{tab:IHS_fluc_statistics2_V2}, \ref{tab:IHS_fluc_statistics2_V1} (IHS) and \ref{tab:LISM_fluc_statistics2_V1} (LISM). It should be noted that  for a fluctuating field with constant magnitude and isotropic angle distribution, C=0.33. 

 These tables also report information on other fluctuations properties such as the average turbulence intensity
 \begin{equation}
I=\left\langle\left|\frac{\delta B}{B_0}\right|\right\rangle,\ \ I_j=\left\langle\left|\frac{\delta B_j}{B_0}\right|\right\rangle, \ \ j=\{\parallel,\perp_1,\perp_2\},
 \end{equation}
 the maximum-variance fluctuation amplitude ($\delta B_{mv}$) and direction with respect to $B_0$ ($\theta_{mv}$), spectral breaks and spectral indexes, power-law exponents for the structure functions.
 
In fact,  we performed a multi-scale analysis of the magnetic field increments via computation of the structure functions $S_p(f)$, a classical and powerful statistical tool to investigate the departure from self-similarity and the intermittent behavior of turbulence   \citep{monin_book, frisch1995, politano1995, politano1998a}.
\begin{equation}\label{eq:strucfun1}
S_{p,j}(\tau)=\left\langle |B_j(t)-B_j(t+\tau)|^p \right\rangle, \ \ j=\{\parallel,\perp_1,\perp_2\}.
\end{equation}
We used the absolute values in this definition for better convergence of statistics for the odd moments.
The computation of $S_{p,j}$ from discrete data is nontrivial for  \textit{Voyager} data sets due to the amount and distribution of missing data. For this computation, we do not interpolate data and compute the statistics for the ensemble of available differences: 
\begin{gather}\label{eq:strucfun2}
S_{p,j}(\tau_k)=\frac{1}{N(\tau_k)}\sum\limits_{i=1}^{N(\tau_k)} |B_j(t_i)-B_j(t_i+\tau_k)|^p,\\ \tau_k=k\cdot\Delta t_s\ \   k=1,\dots,n \nonumber
 \end{gather}
where $\Delta t_s$ is the data resolution and $n$ the total number of points of the data set (we used both 48 s data and 1824 s averages). The counter $N(\tau_j)$ of decreases with $\tau$ and also depends on the distribution of missing data. 
The amount of missing data is between 55\% (\textbf{A1}) and 80\% (\textbf{D1}), and the dominant periodicity of data gaps is 43000 s $\pm$ 2000 s ($f_{gap}=2.3\times 10^{-5}\pm 10^{-6} $ Hz). From the structure functions, one computes the scale-dependent kurtosis of magnetic field increments, which is an indicator of intermittency:
 \begin{equation}\label{eq:kurtosis}
K(\tau)=\frac{S_4(\tau)}{S_2^2(\tau)}.
 \end{equation}
Structure functions and kurtosis are shown in Figures  \ref{fig:SF_V2_IHS} and \ref{fig:SF_V1_IHS} for the IHS, and \ref{fig:V1_LISM_ene} for the LISM. 
 
The spectra computed from \textit{in situ} single-spacecraft measurements are inevitably 1D-reduced spectra \citep{matthaeus1982b} with frequencies measured in the spacecraft reference frame ($f_\mathrm{SC}$). The Doppler-shift relationship between the spacecraft and the plasma reference frame ($f_\mathrm{PL}$) reads
 \begin{equation}
f_\mathrm{SC}=f_\mathrm{PL}+(2\pi)^{-1} \boldsymbol{k}\boldsymbol{\cdot} \mathbf{V_{rel}},
 \end{equation}
where $\boldsymbol{k}$ is the vector wave number and $\mathbf{V_{rel}}=\mathbf{V_{SW}}-\mathbf{V_{SC}}$ the relative speed between the spacecraft and the plasma flow across it. Notice that for an Alfv{\'e}nic nondispersive large-scale wave
, the maximum value is reached for parallel fluctuations, $f_\mathrm{PL}=\kappa_\parallel V_\mathrm{A}/2\pi$. For dispersive waves instead (as in the kinetic regime), $f_\mathrm{PL}$ is typically a function of higher powers of the wavenumber. If $f_\mathrm{PL}\ll|\boldsymbol{\kappa}\boldsymbol{\cdot}\mathbf{V_{rel}}|/2\pi$,  
(Taylor's hypothesis, \cite{taylor1938}, see also the discussion in \cite{howes2014}), frequencies measured at the spacecraft can be converted into wavenumbers in the direction of the relative wind flow. In the solar wind, this condition is satisfied on large scales as the flow is super-Alfv\'enic and the spacecraft is slow compared with the wind. This condition typically holds for the solar wind upstream of the TS, where $V_\mathrm{SC}/V_\mathrm{SW}\lesssim0.05$. In the IHS, the situation differs quite at V1 and V2. At V2, in fact,  $V_\mathrm{SC}/V_\mathrm{SW}\approx0.1$ and $V_\mathrm{A}/V_\mathrm{SW}\approx 0.3$. The B-V angle is nearly equal to $\pi/2$, and the Taylor's approximation might be used to obtain perpendicular wavenumbers, $\kappa_\perp\approx 2\pi f_\mathrm{SC2}/V_\mathrm{SW}$. We report the wavenumber value in all V2 figures. It should be reminded, however,  that reduced spectra always contain contribution from all vector wavenumbers. 

At V1, given that the spacecraft was in the slow-wind region, the Taylor's approximation does not hold.

While performing the turbulence analysis of solar wind fluctuations, it is important to consider the causality condition. In fact, a fluctuation (or ``eddy'') with typical velocity scale $\delta v$ and size $\ell$,  experiences one ``eddy turnover'' in a period $t\sim \pi \ell/\delta v$. During this period, the eddy is convected by the wind by a distance equal to $d=V_\mathrm{SW} t$. Assuming the frozen-flow approximation, this fluctuation would be detected by the spacecraft at a frequency $f_\mathrm{e}\approx \kappa V_\mathrm{SW}/2\pi\approx V_\mathrm{SW}/\ell\approx \pi V_\mathrm{SW} /(\delta v t)\approx \pi V_\mathrm{SW}^2/(\delta v\ d)$. Reminding that $d$ is the distance traveled by the eddy from its origin to the spacecraft, one gets $f_\mathrm{e}\approx \pi V_\mathrm{SW}^2/[\delta v\ (r_\mathrm{SC}-r_\mathrm{source})]$. Eventually,  considering $\delta v \approx V_\mathrm{A}$, the following expression is obtained
 \begin{equation}\label{eq:1_eddy_freq}
 f_\mathrm{e}\approx \frac{\pi V_\mathrm{SW}^2}{V_\mathrm{A}(r_\mathrm{SC}-r_\mathrm{source})}.
 \end{equation} 
 This means that the frequencies less than $f_e$ in the spacecraft reference frame correspond to structures which did not  experience yet  one eddy turnover or, equivalently, waves that do not satisfy the causality condition, since they would have traveled from a further distance than their source point.
 In the solar wind upstream of the TS, the Sun can be clearly considered as the source point. In this case, fluctuations with $f_\mathrm{SC}\lesssim f_e$ would be older than the age of plasma. Fluctuations with $f_\mathrm{SC}\gtrsim f_e$ instead can be considered ``active'' fluctuations, meaning that they may be part of a turbulent energy cascade. 

In this study, we computed $f_e$ at V2 only, and considered the termination shock as the source location. 

Due to the differences  highlighted above between the plasma flow at the two spacecraft and their spatial and temporal separation, magnetic field fluctuations in the IHS at V2 and V1 will be discussed in two separate subsections, \S\ref{sec:V2_IHS} and  \S\ref{sec:V1_IHS}, respectively.  

\subsection{IHS analysis of Voyager 2 data}\label{sec:V2_IHS}
\begin{figure}\centering
\includegraphics[width=0.80\columnwidth]{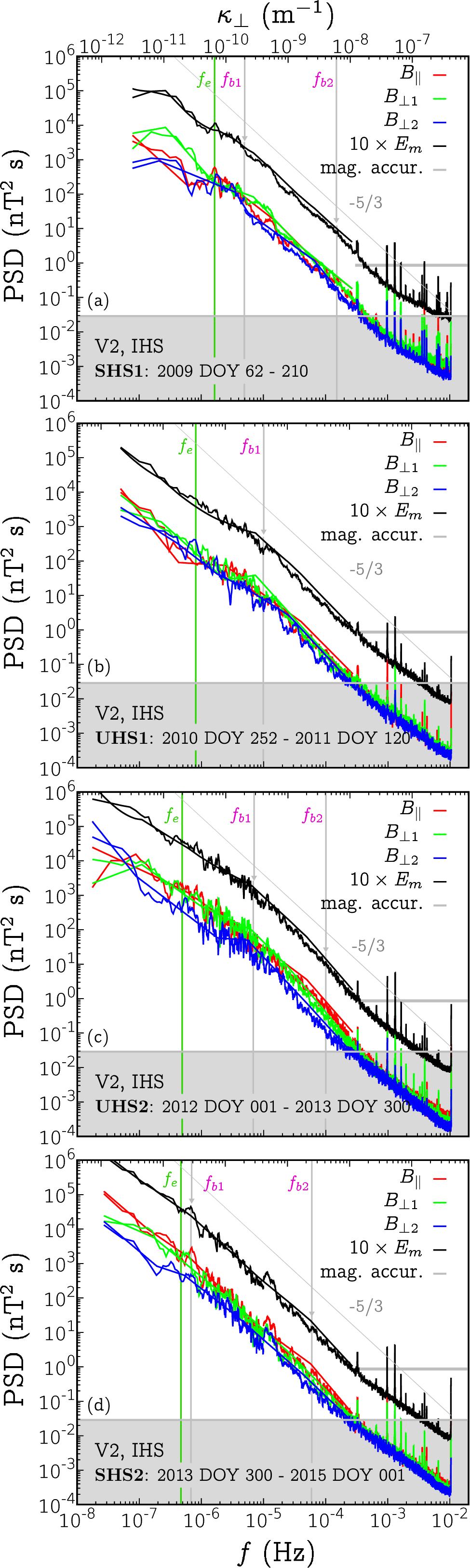}\vskip-10pt
 	\caption{Power spectral density of magnetic field fluctuations at  \textit{Voyager 2} in the IHS.  For clarity, the trace ($E_m$) has been magnified by a factor of 10. Details about the methods for spectral estimation are given in Appendix \ref{sec:App_A}.\label{fig:V2_IHS_ene}}
 \end{figure}

Figure \ref{fig:V2_IHS_ene} shows magnetic field power spectra of the four selected periods of V2 data. Average plasma parameters are summarized in Table \ref{tab:V2_IHS}. Information about the average fluctuating energy,  compressibility and strength of fluctuations is reported in Table \ref{tab:IHS_fluc_statistics2_V2}, together with the frequency of spectral breaks and power-law exponents. Different regimes are identified. 
\begin{deluxetable}{lccccBlcccc}[]
	\tablecaption{Averaged quantities at \textit{Voyager 2} in the IHS. Plasma quantities are computed from PLS data available in the NASA COHO web site. The table reports $r_\mathrm{SC}$, the Sun-V1 distance; $V_0=(V_{0_R}^2+V_{0_T}^2+V_{0_N}^2)^{1/2}$ and $B_0$, the magnitudes of the average velocity and magnetic field, respectively; $B$, the average magnetic field strength. For the thermal protons: $f_\mathrm{cp}$, the cyclotron frequency; $n_\mathrm{p}$, the average density; $T_\mathrm{p}$, average temperature; $V_\mathrm{A}$, the Alfv{\'e}n velocity ; $\beta_\mathrm{p}=\langle 2\mu_0 n_\mathrm{p} k_\mathrm{B}T_\mathrm{p}/B^2\rangle$, the beta of thermal ions; $\beta_\mathrm{p\ 1keV}$, the beta of 1-keV PUIs ($n_{1keV}\approx0.2~n$); $r_\mathrm{ip}$, the ion inertial radius; $r_\mathrm{cp}$, the gyroradius.   $r_\mathrm{cp\ 1keV}$ is the ion gyroradius of a 1-keV pickup proton.  Via the Taylor's approximation, frequencies are also converted in the spacecraft reference frame $f_\mathrm{SC}\approx V_\mathrm{SW}/(2r)$. Eventually, the 1-eddy-turnover frequency $f_e$ is shown (Eq. \ref{eq:1_eddy_freq}). \label{tab:V2_IHS}}
	\tablecolumns{5}
	\tabletypesize{\scriptsize}
	\tablewidth{0pt}
	\tablehead{
		\colhead{\textit{Voyager 2}} &	
		\colhead{\textbf{SHS1}} &
		\colhead{\textbf{UHS1}} &
		\colhead{\textbf{UHS2}} & 
		\colhead{\textbf{SHS2}} 
	}
	\startdata
	$  r_\mathrm{SC} \ \mathrm{(AU)}$         &  89.0 &  94.1 & 100.2  & 104.9 \\
	$  V_0 \ \mathrm{(km\ s^{-1})}$           &  157 &  151 & 154  & 153 \\
	$  B_0 \ \mathrm{(nT)}$                   & 0.062  & 0.072  & 0.090 & 0.030   \\
	$  B \ \mathrm{(nT)}$                   &  0.096 &  0.086 & 0.128  &  0.100   \\
	$  n_p\ \mathrm{(cm^{-3})}$               & 1.1$\times10^{-3}$ &  1.0$\times10^{-3}$ & 1.9$\times10^{-3}$ &   1.8$\times10^{-3}$  \\
	$  T_p\ \mathrm{(10^{4}K)}$               & 6.35   &   4.47 &   5.63 &   5.17     \\
	$  V_\mathrm{A}\ \mathrm{(km\ s^{-1})}$   & 63.4   &  59.2  & 63.9  &  51.6     \\
	$  \beta_\mathrm{p}$                          & 0.54   & 0.46  &  0.54 & 1.00     \\ 
	$  \beta_\mathrm{p~1keV}$                      & 19.6   & 24.3 &  22.2 & 45.0     \\ \hline
	$ r_\mathrm{ip} \mathrm{(km)}$                    & 6852   & 7186   &  5214  & 5357   \\
	$ r_\mathrm{cp}\ \mathrm{(km)}$                   & 3502   &   3301 &   2485 &  3038  \\ 	
	$ r_\mathrm{cp\ 1keV}\ \mathrm{(km)}$             &   53000 &  53165   & 35695  & 45480   \\	\hline
	$ f_\mathrm{ip\ SC}\ \mathrm{(mHz)}$              & 11.5  & 10.5 & 14.8  & 14.3 \\  
	$ f_\mathrm{cp}\ \mathrm{(mHz)}$                  & 1.47 & 1.31 & 1.95 & 1.15 \\
	$ f_\mathrm{cp\ SC}\ \mathrm{(mHz)}$              & 22.4 & 22.8 & 30.9 & 25.2  \\ 
	$ f_\mathrm{cp\ 1keV\ SC}\ \mathrm{(mHz)}$       & 1.65  & 1.42  & 2.16   & 1.68 \\  
	$ f_\mathrm{e}\ \mathrm{(Hz)}$                    & 1.6$\times10^{-6}$ & 8$\times10^{-7}$ & 4$\times10^{-7}$ &  4.5$\times10^{-7}$\\ \hline
	\enddata
\end{deluxetable}

Let us start the discussion with the high-frequency range ($10^{-3}\lesssim f<10^{-2}$ Hz). In principle, 48-s data could allow us to investigate the beginning of the \textit{kinetic regime}, as the ion cyclotron frequency in the IHS is of the order of mHz (see \cite{smith2006a,alexandrova2008b,alexandrova2012}, and \cite{schekochihin2009} for a detailed review). Unfortunately, however, noise affects the data, as explained in \S\ref{sec:data}. Thus, all PSD figures contain a gray band at a power level of $P_\mathrm{noise}=0.029\ \mathrm{nT^2s}$, corresponding to a white noise of 0.03 nT amplitude (the actual distribution of the noise is unknown, this is a conservative threshold). The band is set at $P_\mathrm{noise}=0.086\ \mathrm{nT^2s}$ for $E_\mathrm{m}(f)$. The ``noisy region'' includes frequencies $f\gtrsim 5\times 10^{-4}$ Hz. Here, a spectral flattening \edit1{towards a -1 spectral slope is observed, together with some instrumental-related spikes, harmonics of the sampling rate. Note, however, that the spectral profiles do not correspond to a white noise, and that} in the last frequency decade the spectra show definite trends and retain some information on the anisotropy. Moreover, these trends are not strictly identical among data sets (see for instance the flattening and consecutive steepening at V1 during \textbf{A1} in Figure \ref{fig:V1_IHS_ene}(a)).  \edit1{Taking as an example \textbf{SHS1}, the flattening starts around $f>10^{-3}$ Hz, where $P<4\times10^{-3}$ nT$^2$s. Tests show that this may be due to a white noise with amplitude 0.005 nT and standard deviation 0.003 nT \citep{gallana2016}. This seems suggesting that the actual noise level is below estimates, at least during some periods, and physical results may still be detectable (note that a similar issue occurred for Voyager velocity measurements in \cite{roberts1987a}). Thus, we show PSDs for the full range of frequencies up to the Nyquist's for 48 s resolution. The uncertainty bands reported here should be considered as upper bounds for the spectral region which may be affected by the noise}. 

Moreover, the cyclotron frequency of low-energy PUIs (1 keV) falls within this range, as shown in Table \ref{tab:V2_IHS}. It would be interesting to investigate the effect of PUIs in mediating turbulence and driving kinetic waves, which can affect the high-frequency part of the inertial regime, as shown in \cite{smith2006,cannon2014a,cannon2014b} and  \cite{aggarwal2016} for the solar wind from 1 to 6 AU. In fact, in the IHS the pickup-ion effect is expected to be considerable, since the density of 1-5 keV pickup protons is about 20\% of the thermal protons density \cite{zank2010}. 

Going back to larger scales, most spectra show the presence of a low-frequency regime ($f\lesssim 10^{-5}$ Hz) which we interpret as the \textit{Energy-Injection range} (EI), a reservoir of energy for turbulence. Here, the magnetic energy decays as $1/f$. In particular,  the spectral index $\alpha_1$ falls between 0.7 and 1.3 for all the components. We computed the spectral index by linear regression in the log-log space. The uncertainty due to the fit is always much smaller than that related to the different spectral estimation techniques. Errors are shown in Table \ref{tab:IHS_fluc_statistics2_V2}, they are usually larger in the EI range. 

Interestingly, a spectral break ($f_{b1}$) characterizes the end of the EI range and the beginning of a steeper cascade which will be interpreted as the \textit{Inertial Cascade} (IC) of turbulence. 
The spectral break  between the EI and the IC regimes and the large-scale $\sim1/f$ power law are known to exist in the solar wind upstream the termination shock (first observation is by \cite{tu1984}). \edit1{\cite{roberts2010} investigated such regime with focus on the effect of radial distance and cross-helicity, for both fast, Alfv{\'e}nic, and slow, less Alfv{\'e}nic, regions from 0.3 to 5 AU. In Alfv{\'e}nic regions out to 5 AU, the Ulysses study showed that $f_{b1}$ is between $10^{-5}$ and $10^{-4}$ Hz (\cite{bruno2009,roberts2010}). In non-Alfv{\'e}nic regions, the exact location less clearly determined, and still a current topic of interest  \citep{bruno2018egu}.}  


Interpretations for the nature of this regime include: the superposition of uncorrelated  samples of solar surface turbulence having log-normal distributions of correlation lengths \citep{matthaeus1986} which can determine 1/f energy decay  \citep{montroll1982};  the reflection of primarily outward-traveling Alfv{\'e}n waves in presence of large-scale inhomogeneities \citep{velli1989,perez2013,tenerani2017}. 

In the IHS, a $1/f$ energy decay was observed at V1 during 2009 by \cite{burlaga2012} via multi-fractal analysis. Their observations were limited to the range $f\in[10^{-7},  10^{-5}]$ Hz, which did not allow them to investigate the existence and location of spectral breaks in the IHS, shown here for the first time (the only published power spectra in the IHS are shown in  \citet{burlaga2010,burlaga2012,burlaga2014}).

Figure \ref{fig:V1_IHS_ene} and Table \ref{tab:IHS_fluc_statistics2_V2} show that at V2 the break frequency $f_{b1}$ observed in $E_m(f)$ (black curves)  is about $5\times 10^{-5}$ Hz for \textbf{SHS1}, corresponding to a spatial scale  $\ell_{b1}\approx 0.2$ AU along the wind direction.  Considering the age of fluctuations born at the Sun, one obtains from Eq. (\ref{eq:1_eddy_freq}) a cutoff frequency of $10^{-8}$ Hz. The actual break is instead more consistent with $f_e$ of fluctuations generated at the TS (or affected by it). Moreover, it is not physically reasonable to consider turbulent structures with size greater than the outer scale of the system. The IHS width observed by V1 is around 27 AU, corresponding to $\kappa\approx 1.5\times 10^{-12}\ \mathrm{m}^{-1}$ (note the wavenumber axis $\kappa\sim\kappa_\perp$ in all V2 plots). Moreover, the Sun's rotation acts as a forcing with $f_\mathrm{sun}\approx 4\times10^{-7}$ Hz. The sector spacing should be around 2 AU after TS, decreasing as the HP is approached, even though the canonical sector structure is no longer recognizable beyond 10 or 20 AU. 

It seems that $f_{b1}$ increases in the unipolar periods \textbf{UHS1} and \textbf{UHS2} to values around $10^{-5}$ Hz ($\ell_{b1}\approx 0.1$ AU), while it decreases again to $f_{b1}\approx7\times10^{-7}$ Hz ($\ell_{b1}\approx1.5$ AU) in the sectored interval \textbf{SHS2}, where the break is actually very weak. Note also that the break location differs for the $\delta B_\parallel$, $\delta B_{\perp1}$ and $\delta B_{\perp2}$ components, represented by the red, green  and blue curves, respectively, in Figure \ref{fig:V2_IHS_ene}. In fact, the break of the B-perpendicular fluctuations occurs at a higher frequency with respect to the B-parallel one, by a factor between 2 and 7 for all periods except the last one. \edit1{We can't be sure due to noisy plasma data, but unipolar regions are at least initially much more Alfv{\'e}nic (and thus have higher $f_{b1}$) than sector regions, and this might account for the differences seen here between the sectored and unipolar regions.}

The high slope in the power spectra at $f\gtrsim f_{b1}$ suggests that a \textit{turbulent inertial cascade} is ongoing.  Moreover, in all intervals except \textbf{UHS1} a second spectral knee is observed at $f_{b2}\approx10^{-4}$ Hz ($\ell_{b2}\approx0.01$ AU). It is particularly visible in the $B_\parallel$ component, while it is weaker and not always observed in $B_{\perp}$. The spectral slope between the two breaks, $\alpha_2$, is about -1.6 for \textbf{SHS1} and \textbf{SHS2}, \textcolor{black}{close to the Kolmogorov' s  -5/3 value \citep{kolmogorov1941a} and compatible with the model of \cite{goldreich1995,goldreich1997}  for the k-perpendicular cascade of critically balanced turbulence. The latter model, however, is not quite adequate, since it ignores the compressibility, which plays a fundamental role in heliosheath and interstellar turbulence.}

During unipolar periods  the magnetic energy decays slightly faster and the index is around -1.75. Note that, in general, the perpendicular components contribute to the steepening of the spectra, and that the \textbf{UHS2}  spectrum changes slope in rather a continuous way across frequencies. Beyond $f_{b2}$, the slope increases to $\alpha_3\approx-2$. The cyclotron frequency of thermal protons, $f_\mathrm{cp}$,  is around 2 mHz (see Table \ref{tab:V2_IHS}). Moreover, structures with size of the Larmor radius convected across the spacecraft may affect the power spectra at spacecraft-frame frequencies close to $f_\mathrm{cp\ SC}\approx 10^{-2}$ Hz. Structures of size comparable with the ion inertial radius, similarly, should be detected at $f_\mathrm{ip\ SC}\approx2.5\times10^{-2}$ Hz. In all likelihood,  the second  break $f_{b2}$ is still within the MHD inertial range. However, we suggest that gyroradii of 1 keV PUIs may affect the turbulence at  $f_\mathrm{cp\ 1keV\ SC}\approx10^{-3}$ Hz (this frequency shifts to lower values as $E(eV)^{-1/2}$). We see that in the range  $f_{b2}\lesssim f \lesssim 10^{-3}$ Hz a reduction of compressibility and intermittency takes place. 

To simplify comparisons and prevent misunderstandings, we emphasize that our $P[B_{\perp 1}]$ is sometimes referred to as the ``perpendicular spectrum'',  whereas $P[B_{\perp 2}]$ corresponds to the ``quasi-parallel'' spectrum \citep{matthaeus1990, bieber1996}. The ratio of these two spectra is equal to 1 in case of pure slab turbulence and 1.67 in the case of pure 2D turbulence. The slab/2D model is not descriptive of the IHS or LISM turbulence, as it ignores compressible fluctuations. 

\begin{figure}[]
		\plotone{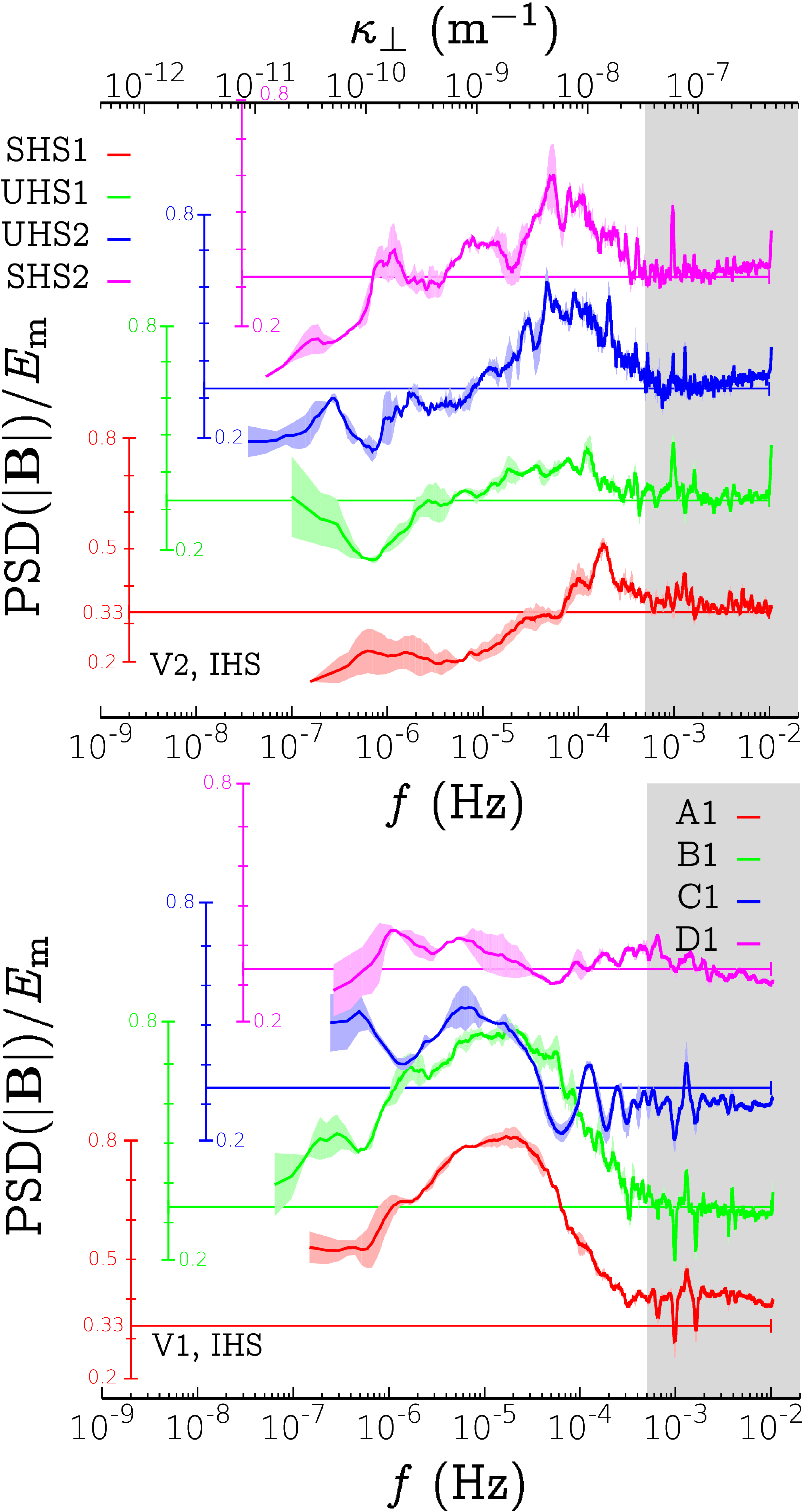}\vspace{-3pt}
	\caption{Spectral compressibility in the IHS at \textit{Voyager 2} ({top panel}) and \textit{Voyager 1} (bottom). The colored areas show the variability due to the methods we used for the computation of the spectrum (see Appendix \ref{sec:App_A}). Average values computed through Eq. \ref{eq:compressibility} are shown in Tables \ref{tab:IHS_fluc_statistics2_V2} and  \ref{tab:IHS_fluc_statistics2_V1} for V2 and V1, respectively. \label{fig:spec_compressibility}}
\end{figure}

The spectral compressibility is shown in Figure \ref{fig:spec_compressibility} (top panel), \edit1{while the variance anisotropy is shown in Appendix \ref{sec:App_B} (Figure \ref{fig:compress_anis_IHS_V2}). It is seen that the magnetic field fluctuations are primarily transverse in the EI range. At the beginning of the IC regime  $P[B_\perp]/P[B_\parallel]\approx2$, meaning that $\delta B_\parallel$ accounts for $\sim$30\% of the energy, and it approaches the unity at $f_{b2}$, where the maximum of compressibility is indeed observed ($\sim$50\% of the energy in $\delta B_\parallel$). Such values of compressibility are relatively large if compared to near-Earth solar wind. The present IHS observations seem consistent with those by \citet{smith2006} from ACE observations at 1 AU, provided we take into account for the large $\beta_p$ due to the PUI population.}

\begin{figure*}\centering
\includegraphics[width=0.77\textwidth]{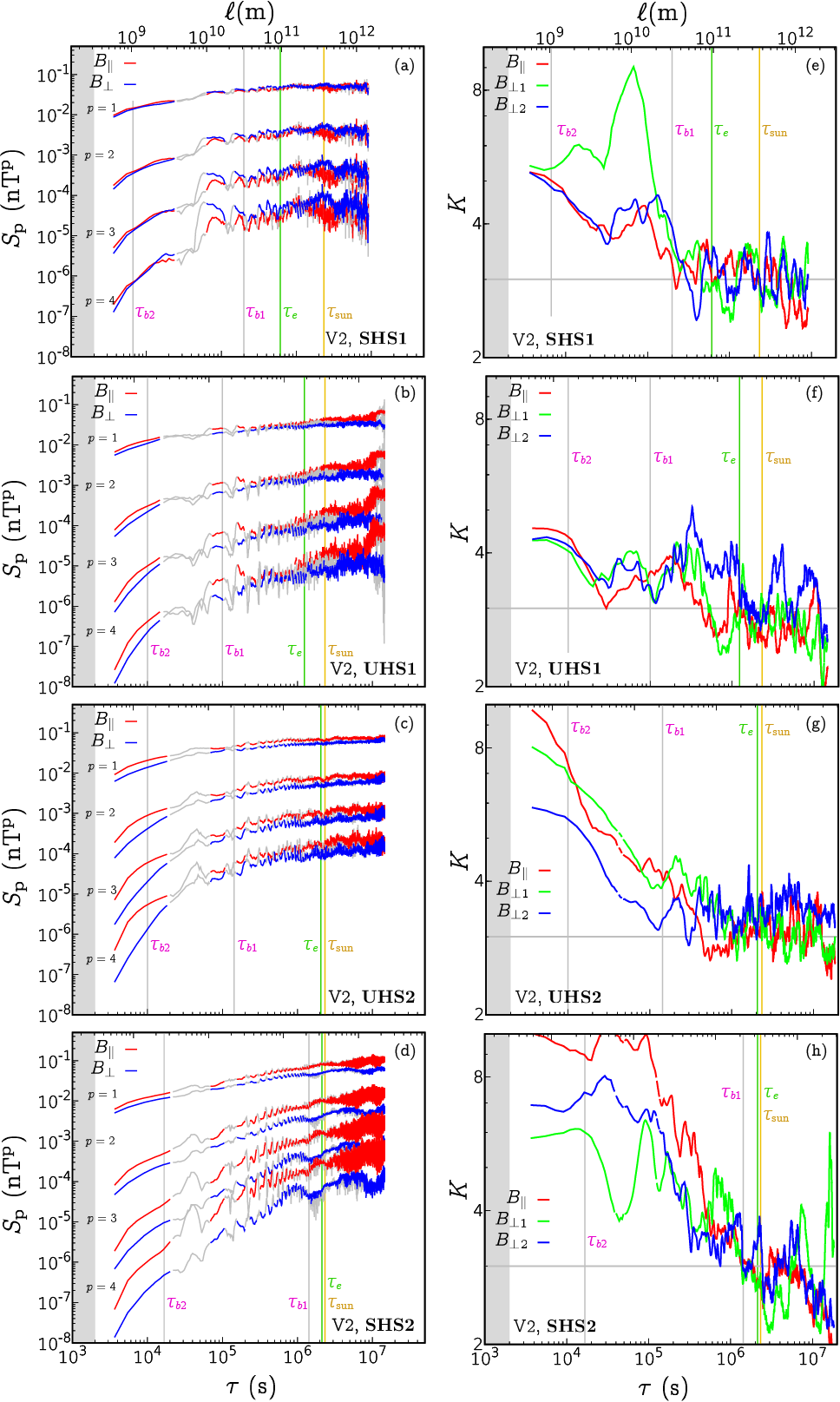}\vspace{-10pt}
	\caption{Structure functions and intermittency at \textit{Voyager 2} in the IHS. Left panels (a-d): structure functions of B-parallel fluctuations, $S_{p,\parallel}$ (red curves) and B-perpendicular fluctuations $S_{p,\perp}=(S_{p,\perp 1}+S_{p,\perp 2})/2$ (blue curves). The time scales of spectral breaks (see Figure \ref{fig:V2_IHS_ene}), $\tau_{b1}$ and $\tau_{b2}$, are indicated with gray vertical lines. Time scales corresponding to the solar rotation, $\tau_\mathrm{sun}$, and $\tau_e=1/f_e$  are also shown. Right panels (e-h): kurtosis of magnetic field increments, obtained via Eq. \ref{eq:kurtosis}.\label{fig:SF_V2_IHS}}
\end{figure*}

The existence of a turbulent inertial range is further investigated by the analysis of structure functions of temporal increments of the magnetic field, $S_p(\tau)$, for $p={1,2,3,4}$ (Eq. \ref{eq:strucfun1}, \ref{eq:strucfun2}). Results are shown in Figure \ref{fig:SF_V2_IHS} in terms of B-parallel and B-perpendicular structure functions (left column) and kurtosis (right). Similarly to the neutral-fluid turbulence, under the assumption of homogeneity and isotropy, the inertial range of MHD turbulence is defined as the range of scales where the  third-order longitudinal structure function displays a linear dependence for the Els{\"a}sser field,   $S_3\sim \tau$. More generally,  a linear dependence occurs for the longitudinal flux of energy, a result known as the Yaglom's four-thirds law \citep{monin_book}, extended to the MHD case by \cite{politano1998b}. This quantity provides information on the rate of dissipation of the turbulent energy and related plasma heating \citep{sorriso2007,carbone2009,hadid2018,sorriso2018}.

In the Kolmogorov's description of (non-intermittent) isotropic and homogeneous turbulence in fluids, $S_p(\tau)\sim \tau^{p/3}$ \citep{kolmogorov1941a}. The presence of intermittency makes the actual scaling exponents of both the velocity and the magnetic fields deviate from the linear trend. For models and observations, the reader is referred to \citet{she1994,politano1995,politano1998b,bruno2013}, and \citet{muller2000}. Table 1 in \citet{politano1998a} shows a comparison of measured exponents with different theoretical predictions.

The structure functions of magnetic field in the IHS at V2 display a power-law behavior in the range of time scales which approximately corresponds to the frequency range between the spectral breaks identified in Figure \ref{fig:V2_IHS_ene}. In the left panels of Figure \ref{fig:SF_V2_IHS}, the red curves stands for $S_{p,\parallel}$ and the blue curves stands for $(S_{p,\perp 1}+S_{p,\perp 2})/2$ (Eq. \ref{eq:strucfun1}). The structure functions are computed from Eq. \ref{eq:strucfun2} and show oscillations related to data gaps. The counter $N(\tau)$ is indeed a oscillating function of  $\tau$, and on average it decreases linearly with $\tau$ (Eq. \ref{eq:strucfun2}). When it is less than a certain threshold (specifically, when $N(\tau)< 0.25 \max[N(\tau')],\ \tau'\in[\tau-48h,\ \tau+48h]$), the color of curves is switched to gray. These points were not used to compute the scaling exponents $\zeta_p$. 

As seen from the line curvature, each scaling exponent changes continuously across scales. However, the energy-injection range is easily identified, as well as the effect of the solar rotation. \edit1{Both the scaling exponents and relative exponents ($\zeta_p^\mathrm{ess}$) are reported in Table \ref{tab:IHS_fluc_statistics2_V2}. Fits for $\zeta_p$ have been computed in the range of $\tau$ between $\tau_{b1}$ and $\tau_{b2}$, shown in the pictures. Relative exponents are computed by fitting $S_p/S_3$. The \textit{Extended Self-Similarity} principle \cite{benzi1993} is well verified, as $S_p/S_3$ show a defined power-law trend well beyond the inertial range, which allows the computation of exponents with good accuracy.}   

Relative exponents of parallel and perpendicular fluctuations are similar, and \textcolor{black}{appear to be closer to the values typical of plasma velocity rather than magnetic field \citep{politano1995}}. These values are consistent with the presence of inertial-range intermittency, which is confirmed by the familiar profiles of kurtosis shown in the  panels (e)-(h) of Figure \ref{fig:SF_V2_IHS}. In fact, the intermittency increases with frequency and starts to increase at the beginning of the IC range, approximately at $f_{b1}$. The EI range is characterized by Gaussian values ($K\approx3$), or even sub-Gaussian. In the inertial-cascade range, $K(\tau)$ rises up to 10. It seems that a damping of such growth can occur at some point within the IC regime, see e. g. the period \textbf{SHS2}. This is also observed at V1 (Figure \ref{fig:SF_V1_IHS}, left panels). The evolution of spectral compressibility seen in Figure \ref{fig:spec_compressibility} (top),  suggests the existence of a relationship  between intermittency and compressibility, as has been shown by \cite{alexandrova2008b} in the kinetic regime. The decrease of compressibility and intermittency in the high-frequency range deserves further investigation, at present we could either interpret it as (i) an effect of data noise or (ii) physical reasons, as observed in \citet{sorriso2017} at 1 AU. Again, the effect of the pickup ion populations should be considered.

It is worth noticing that the intermittency of magnetic fluctuations in the heliosheath was investigated earlier in the framework of a multi-fractal formalism \citep{meneveau1987,frisch1995}, see e.g. \cite{burlaga2006b,macek2012,burlaga2010,burlaga2013b,macek2013,macek2014}. Most of the published analyses, however, are focused on the magnetic field magnitude and consider scales larger than one day.  Notable exceptions are presented by \citet{burlaga2009} and  \citet{burlaga2013b}, who consider the probability distribution functions of the increments of magnetic field magnitude. \citet{burlaga2009} also used 48 s data and provided a description of different magnetic structures observed in the IHS. 
We point out that the low values of multi-fractal index reported, e.~g., by \citet{macek2014} should not be interpreted as a non-intermittent inertial range of turbulence. In fact, their range of scales corresponds to the EI range in the present study.

\begin{deluxetable}{l|cccc}[]
	\tablecaption{Magnetic field fluctuations properties at V2 in the IHS. All quantities are defined in \S \ref{sec:V2_IHS}. Spectral breaks and indexes of perpendicular fluctuations refer to the total power spectrum $\mathrm{P}[B_\perp]=\mathrm{P}[B_{\perp1}]+\mathrm{P}[B_{\perp 2}]$.    \label{tab:IHS_fluc_statistics2_V2}}
	\tabletypesize{\scriptsize}
	\tablecolumns{7}
	\tablewidth{0pt}
	\tablehead{
		\colhead{\textit{Voyager 2}} &	
		\colhead{\textbf{SHS1}} &
		\colhead{\textbf{UHS1}} &
		\colhead{\textbf{UHS2}} & 
		\colhead{\textbf{SHS2}} 
	}
	\startdata 
	$  E_m \ \mathrm{(nT^2)}$           & 6.76$\times10^{-3}$   & 3.42$\times10^{-3}$   &  1.17$\times10^{-2}$ &1.18$\times10^{-2}$    \\
	$C_{2}$                & 0.29& 0.42 & 0.35& 0.43\\
	$C_{2}$                & 0.35& 0.63 & 0.48& 0.64\\
	$I_\parallel$               & 0.50& 0.45  & 0.52& 2.29\\
	$I_{\perp1}$                & 0.62& 0.37 & 0.50& 2.14\\
	$I_{\perp2}$                & 0.44& 0.25 & 0.49& 1.06\\
	$I$                          & 1.04& 0.74  & 1.00& 3.70\\ 
	$\delta B_\mathrm{mv}$         & 0.051& 0.043 & 0.067& 0.088\\ 
	$\theta_{mv}$    & 106$^\circ$&153$^\circ$ & 86$^\circ$&51$^\circ$ \\  \hline\hline
	$f_\mathrm{b1}$ (Hz)                  &5$\times10^{-6}$ & $10^{-5}$ &$7\times10^{-6}$ &$7\times10^{-7}$  \\ 
	$f_\mathrm{b2}$ (Hz)                  & 2$\times10^{-4}$ & \nodata&$10^{-4}$ &7$\times10^{-5}$  \\   
	$\alpha_\mathrm{1}$               &-0.98$\pm$0.10 &-1.10$\pm$0.03&-1.18$\pm$0.05 &-1.25$\pm$0.13  \\
	$\alpha_\mathrm{2}$               &-1.64$\pm$0.01 &-1.78$\pm$0.06&-1.78$\pm$0.13 &-1.58$\pm$0.02  \\ \hline
	$f_\mathrm{b1, \parallel}$ (Hz)        &2$\times10^{-6}$ & 3$\times10^{-6}$ & 2$\times10^{-6}$&7$\times10^{-7}$ \\ 
	$f_\mathrm{b2,\parallel}$ (Hz)        &2$\times10^{-4}$ &$10^{-4}$ & 5$\times10^{-5}$&  6$\times10^{-5}$ \\ 
	$\alpha_\mathrm{1, \parallel}$      & -0.74$\pm$0.10 &-0.97$\pm$0.12 & -0.84$\pm$0.05& -1.15$\pm$0.02 \\
	$\alpha_\mathrm{2, \parallel}$       &-1.62$\pm$0.02& -1.60$\pm$0.10& -1.34$\pm$0.05 & -1.57$\pm$0.03 \\ 
	$\alpha_\mathrm{3, \parallel}$           & -1.97$\pm$0.02 & \dots& -2.08$\pm$0.02 & -2.02$\pm$0.08 \\ \hline
	$f_\mathrm{b1,\perp}$ (Hz)             &5$\times10^{-6}$ & 2$\times10^{-5}$ & 8$\times10^{-6}$&7$\times10^{-7}$\\ 
	$f_\mathrm{b2,\perp}$ (Hz)         &2$\times10^{-4}$ &\nodata & \nodata& 7$\times10^{-5}$ \\   
	$\alpha_\mathrm{1, \perp}$          & -0.80$\pm$0.10 &-1.25$\pm$0.04 & -1.13$\pm$0.03& -1.11$\pm$0.10\\
	$\alpha_\mathrm{2, \perp}$         &-1.64$\pm$0.02 &-1.87$\pm$0.06  & -2.05$\pm$ 0.05& -1.58$\pm$0.02\\
	$\alpha_\mathrm{3,\perp}$          &-1.70$\pm$0.05 & \nodata & \nodata & -1.70$\pm$0.05
	\\     \hline \hline    
	$ \zeta_\mathrm{1, \parallel}$ \textcolor{black}{$ (\zeta_\mathrm{1, \parallel}^{\mathrm{ess}})$}    & 0.33 \textcolor{black}{\bf (0.35)} & 0.30 \textcolor{black}{\bf(0.33)} & 0.28 \textcolor{black}{\bf(0.41)} & 0.40 \textcolor{black}{\bf(0.41)} \\  
	$ \zeta_\mathrm{2, \parallel}$ \textcolor{black}{$ (\zeta_\mathrm{2, \parallel}^{\mathrm{ess}})$}      & 0.67 \textcolor{black}{\bf(0.68)} & 0.57 \textcolor{black}{\bf(0.67)}&  0.48 \textcolor{black}{\bf(0.74)} & 0.75 \textcolor{black}{\bf(0.75)}\\
	$ \zeta_\mathrm{3, \parallel}$ \textcolor{black}{$ (\zeta_\mathrm{3, \parallel}^{\mathrm{ess}})$}    & 1.02 (1) & 0.82 (1) &  0.67 (1)& 1.05 (1)\\     
	$ \zeta_\mathrm{4, \parallel}$ \textcolor{black}{$ (\zeta_\mathrm{4, \parallel}^{\mathrm{ess}})$}          & 1.38 \textcolor{black}{\bf(1.30)} &  1.04 \textcolor{black}{\bf(1.31)} &  0.89 \textcolor{black}{\bf(1.20)}& 1.31 \textcolor{black}{\bf(1.18)}\\       \hline         
	$ \zeta_\mathrm{1, \perp}$   \textcolor{black}{$ (\zeta_\mathrm{1, \perp}^{\mathrm{ess}})$}    & 0.40 \textcolor{black}{\bf(0.37)} & 0.25 \textcolor{black}{\bf(0.34)} &  0.37 \textcolor{black}{\bf(0.38)}& 0.33 \textcolor{black}{\bf(0.36)}\\                  
	$ \zeta_\mathrm{2, \perp}$  \textcolor{black}{$ (\zeta_\mathrm{2, \perp}^{\mathrm{ess}})$}     & 0.82 \textcolor{black}{\bf(0.71)} & 0.45 \textcolor{black}{\bf(0.67)}&  0.62 \textcolor{black}{\bf(0.71)}& 0.62 \textcolor{black}{\bf(0.70)}\\          
	$ \zeta_\mathrm{3, \perp}$   \textcolor{black}{$ (\zeta_\mathrm{3, \perp}^{\mathrm{ess}})$}    & 1.25 (1)& 0.64 (1) & 0.81  (1)& 0.87 (1)\\          
	$ \zeta_\mathrm{4, \perp}$  \textcolor{black}{$ (\zeta_\mathrm{4, \perp}^{\mathrm{ess}})$}     & 1.69 \textcolor{black}{\bf(1.24)} & 0.82 \textcolor{black}{\bf(1.30)} &  0.98 \textcolor{black}{\bf(1.26)} & 1.10 \textcolor{black}{\bf(1.25)}\\   \hline\hline               
	\enddata
\end{deluxetable}

\FloatBarrier
\subsection{IHS analysis of Voyager 1 data}\label{sec:V1_IHS}
\begin{figure}\centering
\includegraphics[width=0.83\columnwidth]{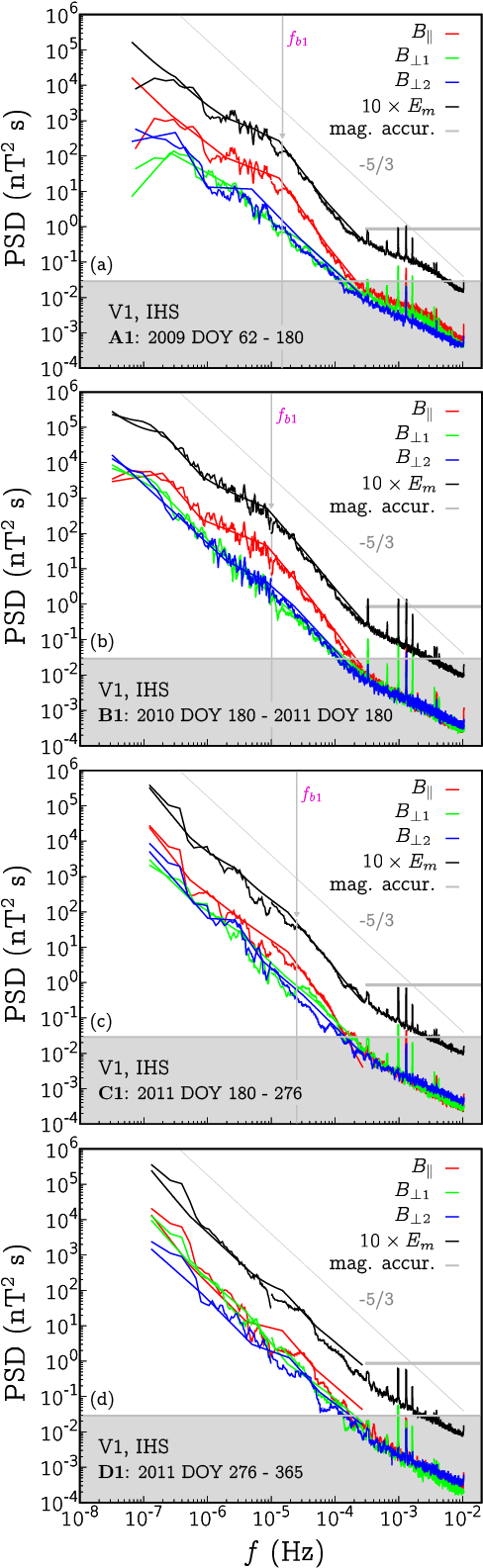}
	\caption{Power spectral density of magnetic field fluctuations at  \textit{Voyager 1} in the IHS.\label{fig:V1_IHS_ene}}
\end{figure}

\begin{figure*}\centering
\includegraphics[width=0.8\textwidth]{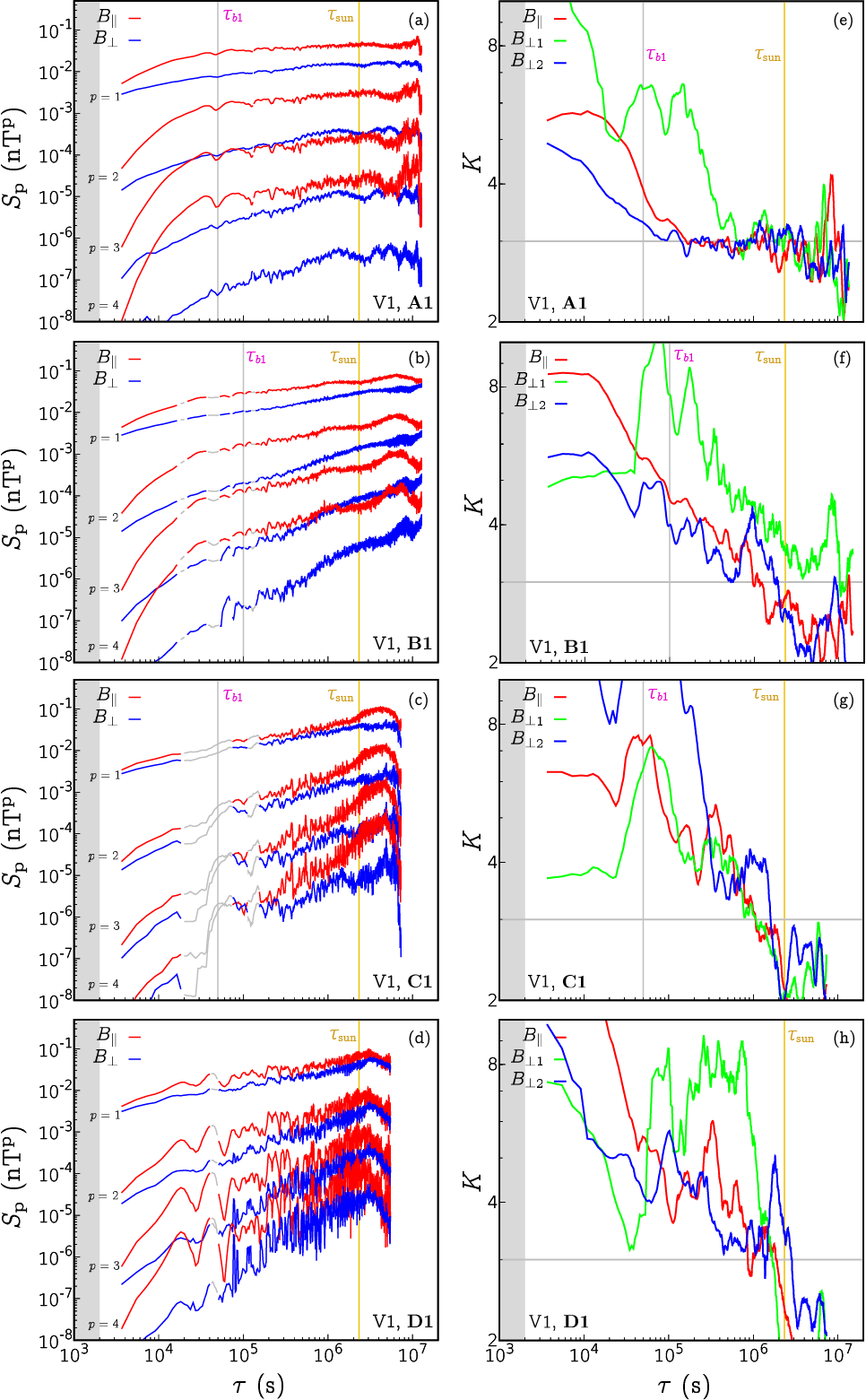}
	\caption{Structure functions (panels a-d) and kurtosis (e-h) at \textit{Voyager 1} in the IHS (see caption of Figure \ref{fig:SF_V2_IHS}).\label{fig:SF_V1_IHS}}
\end{figure*}

Figure \ref{fig:V1_IHS_ene} shows power spectra for  the V1 intervals \textbf{A1}, \textbf{B1}, \textbf{C1} and \textbf{D1}. Average quantities are reported in Table \ref{tab:V1_IHS}, while the fluctuation statistics, spectral breaks and slopes, and structure-function exponents are shown in Table \ref{tab:IHS_fluc_statistics2_V1}. Spectral compressibility is shown in Figure \ref{fig:spec_compressibility} (bottom panel), and high-order statistics are shown in Figure \ref{fig:SF_V1_IHS}.

 At V1, the intensity of magnetic fluctuations is in general smaller compared to V2. The contribution to the fluctuating energy is largely due to the $\delta B_\parallel$ components, which results in much larger values of compressibility ($C_1\approx0.6$), especially during 2009-2011.5. 

Since the Taylor's hypothesis does not hold at V1, we do not convert spacecraft frequencies to wavenumbers and all figures show the spacecraft-frame frequency axis only. In the early periods \textbf{A1} and \textbf{B1}, the spectra show a marked difference between B-parallel and B-perpendicular fluctuations,  $\delta B_\parallel$ being more energetic than $\delta B_\perp$ by a factor as high as five in the central decades of the spectrum.  \edit1{Details on variance anisotropy are given in Figure \ref{fig:compress_anis_IHS_V1} in Appendix \ref{sec:App_B}, where it is shown that $P[B_\perp]/P[B_\parallel]<1$ in  proximity of the spectral break observed at $f_{b1}\approx10^{-5}$ Hz. This corresponds to the presence of compressive modes. In fact, $P[|\mathbf{B}|]/E_m$ reaches the maximum at $f_{b1}$ (see Figure \ref{fig:spec_compressibility}, bottom, and the black curves in left panels of Figure \ref{fig:compress_anis_IHS_V1}).} 
At lower frequencies, the spectral index of the total energy  is $\alpha_1\approx-1.2$ (black curves in Figure \ref{fig:V1_IHS_ene}), while beyond the break a fast steepening occurs, with the slopes as high as $\alpha_2\approx -2.3$. Again, the shape of the spectral trace is mainly due to $\delta B_\parallel$, which displays a rather fast cascade in the range $10^{-5}\lesssim f \lesssim 10^{-4}$ Hz ($\alpha_{2 ,\parallel}\approx-2.5$). The two perpendicular components behave similarly to each other, and on average they experience a Kolmogorov-like spectral decay in the whole range of frequencies, with index between -1.35 and -1.8.  In later periods \textbf{C1} and \textbf{D1}, which are closer to the HP boundary, the spectral break becomes weaker even for $\delta B_\parallel$, so that the discrepancy between the components is significantly reduced (see the bottom panels in Figure \ref{fig:V1_IHS_ene} \edit1{and left panels of Figure \ref{fig:compress_anis_IHS_V1}}). In fact, the level of compressibility decreases to 0.4 during \textbf{D1} at all frequencies. 
\begin{deluxetable}{lccccBlcccc}[t]
	\tablecaption{Averaged quantities at V1 in the IHS. Since the PLS subsystem is not operative, the velocity is derived from LECP and CRS subsystems. Here, we report data  from Figure 1 in \cite{krimigis2011} and  Figure 1 in \cite{richardson2016b}. 
		\label{tab:V1_IHS}}
	\tablecolumns{5}
	\tablewidth{0pt}
	\tablehead{
		\colhead{\textit{Voyager 1}} &	
		\colhead{\textbf{A1}} &
		\colhead{\textbf{B1}} &
		\colhead{\textbf{C1}} & 
		\colhead{\textbf{D1}} 
	}
	\startdata
	$  r_\mathrm{SC} \ \mathrm{(AU)}$      &  109.5 &  115.7 & 117.9  & 118.8 \\
	$  V_0 \ \mathrm{(km\ s^{-1})}$      &  65 &  40 & 40  & 40 \\
	$  B_0 \ \mathrm{(nT)}$           &   0.083 &   0.132 & 0.195  & 0.124   \\
	$  B \ \mathrm{(nT)}$                   &  0.086 &  0.140 & 0.203  &  0.148   \\
	$ f_\mathrm{cp}\ \mathrm{(mHz)}$                   & 1.32 & 2.14 & 3.10 & 0.81 \\ \hline
	\enddata
\end{deluxetable}

At frequencies higher than about $10^{-4}$ Hz, the spectra flatten, \edit1{likely} due to the lower limit of data accuracy. The curved shape at $10^{-3}<f<10^{-2}$ Hz during \textbf{A1}, however, suggests that physical phenomena such as wave-particle interaction or PUI-driven turbulence may still be relevant in the signal. 

The profiles of parallel structure functions (red curves in Figure \ref{fig:SF_V1_IHS}, left column)  show a well-defined change in the power law on the time scales corresponding to the observed spectral break.  The exponents $\zeta_p$ reported in Table \ref{tab:IHS_fluc_statistics2_V1}  have been fitted in the range $\tau\in[10^4\ s, \tau_{b1}]$ for \textbf{A1} and \textbf{B1}, and $\tau\in[10^4, 2\times10^5]$ s for \textbf{C1} and \textbf{D1}. Thus, we show the exponents  of $S_p$ for the $\alpha\approx-2$ part of the power spectra. Relative exponents instead hold for the whole range of frequencies.  Also, at \textit{Voyager 1}, the fluctuations are intermittent (right panels of Figure \ref{fig:SF_V1_IHS}). In addition to previous large-scale analyses \citep[e.g.][]{burlaga2006,macek2014}, we show that the  kurtosis profiles of magnetic increments increase with increasing frequency for all intervals. It should be noted, however, that the growth starts within the range $\tau\in[10^5,10^6]$ s. This was expected for the $B_\perp$ components, but not so for  $B_\parallel$: is seems that the intermittency of the compressible component starts prior to the energy spectral break, at about $f\approx10^{-6}$ Hz, except for the interval \textbf{A1}.  


\begin{deluxetable}{l|cccc}[]
	\tablecaption{Magnetic field fluctuations properties at V1 in the IHS.\label{tab:IHS_fluc_statistics2_V1}}
	\tablecolumns{5}
	\tablewidth{0pt}
	\tablehead{
		\colhead{} & 	\textbf{A1} & \textbf{B1} & \textbf{C1} & \textbf{D1}	}
	\startdata 
	$  E_\mathrm{m} \ \mathrm{(nT^2)}$           & 1.67$\times10^{-3}$   & 5.03$\times10^{-3}$   &  4.50$\times10^{-3}$ &4.05$\times10^{-3}$    \\	
	$C_\mathrm{1}$                 & 0.61& 0.55& 0.62& 0.42 \\
	$C_\mathrm{2}$                 & 0.95& 0.76& 0.74& 0.62 \\
	$I_\parallel$                 & 0.34& 0.40& 0.24& 0.28 \\
	$I_{\perp1}$                 & 0.11& 0.17& 0.10& 0.27 \\
	$I_{\perp2}$                  & 0.13& 0.22& 0.12& 0.13 \\
	$I$                            & 0.42& 0.54& 0.31& 0.46 \\ 
	$\delta B_\mathrm{mv}$         & 0.037& 0.061& 0.069& 0.055 \\ 
	$\theta_{mv}$   & 166$^\circ$&147$^\circ$&141$^\circ$ &42$^\circ$ \\  \hline
	$f_\mathrm{b1}$ (Hz)                  & $2\times10^{-5}$& $10^{-5}$& $2\times10^{-5}$& \nodata  \\  
	$\alpha_\mathrm{1}$               &-1.17$\pm$0.09 &-1.31$\pm$0.15 &-1.60$\pm$0.18 & \nodata \\
	$\alpha_\mathrm{2}$               & -2.25$\pm$0.12& -2.31$\pm$0.04& -2.23$\pm$0.09& -1.72$\pm$0.05\\ \hline
	$f_\mathrm{b1\parallel}$ (Hz)         & $2\times10^{-5}$& $10^{-5}$&$2\times10^{-5}$ & \nodata\\ 
	$\alpha_\mathrm{1\parallel}$       &-1.18$\pm$0.06 & -1.21$\pm$0.07 &  -1.45$\pm0.10$& -1.95$\pm0.10$\\
	$\alpha_\mathrm{2\parallel}$       & -2.65$\pm$0.10 & -2.50$\pm$0.03& -2.52$\pm$0.10&  -1.80$\pm0.05$\\ \hline
	$f_\mathrm{b1\perp}$ (Hz)             & $4\times10^{-6}$ & \nodata& \nodata& \nodata  \\   
	$\alpha_\mathrm{1\perp}$           &-1.35$\pm$0.04 & -1.52$\pm$0.05& -1.57$\pm0.09$&-1.76$\pm0.10$  \\
	$\alpha_\mathrm{2\perp}$           & -1.59$\pm$0.13 & -1.77$\pm$0.06& -1.70$\pm0.1$& -1.80$\pm0.05$ 
	\\     \hline     
	$ \zeta_\mathrm{1, \parallel}$ \textcolor{black}{$ (\zeta_\mathrm{1, \parallel}^{\mathrm{ess}})$}                    & 0.48 \textcolor{black}{\bf(0.38)} & 0.62 \textcolor{black}{\bf(0.36)} & 0.44 \textcolor{black}{\bf(0.39)} & 0.36 \textcolor{black}{\bf(0.33)} \\  
	$ \zeta_\mathrm{2, \parallel}$    \textcolor{black}{$ (\zeta_\mathrm{2, \parallel}^{\mathrm{ess}})$}                 & 0.93 \textcolor{black}{\bf(0.71)} & 1.19 \textcolor{black}{\bf(0.70)}&  0.90 \textcolor{black}{\bf(0.72)} & 0.62 \textcolor{black}{\bf(0.65)}\\
	$ \zeta_\mathrm{3, \parallel}$  \textcolor{black}{$ (\zeta_\mathrm{3, \parallel}^{\mathrm{ess}})$}                   & 1.31 (1) &  1.63 (1) & 1.34 (1)& 0.81 (1)\\     
	$ \zeta_\mathrm{4, \parallel}$   \textcolor{black}{$ (\zeta_\mathrm{4, \parallel}^{\mathrm{ess}})$}                  &  1.64 \textcolor{black}{\bf(1.24)} &  1.99 \textcolor{black}{\bf(1.24)} &  1.72 \textcolor{black}{\bf(1.27)}& 0.98 \textcolor{black}{\bf(1.23)}\\       \hline         
	$ \zeta_\mathrm{1, \perp}$   \textcolor{black}{$ (\zeta_\mathrm{1, \perp}^{\mathrm{ess}})$}                  & 0.34 \textcolor{black}{\bf(0.36)} & 0.41 \textcolor{black}{\bf(0.36)} &  0.43 \textcolor{black}{\bf(0.32)}& 0.36 \textcolor{black}{\bf(0.38)}\\                  
	$ \zeta_\mathrm{2, \perp}$   \textcolor{black}{$ (\zeta_\mathrm{2, \perp}^{\mathrm{ess}})$}                  &  0.67 \textcolor{black}{\bf(0.70)}& 0.81 \textcolor{black}{\bf(0.69)}& 1.02 \textcolor{black}{\bf(0.66)}& 0.74 \textcolor{black}{\bf(0.70)}\\          
	$ \zeta_\mathrm{3, \perp}$  \textcolor{black}{$ (\zeta_\mathrm{3, \perp}^{\mathrm{ess}})$}                   & 0.97 (1)& 1.17 (1)& 0.87 (1)& 1.14 (1)\\          
	$ \zeta_\mathrm{4, \perp}$  \textcolor{black}{$ (\zeta_\mathrm{4, \perp}^{\mathrm{ess}})$}                   &  1.23 \textcolor{black}{\bf(1.24)}&  1.48 \textcolor{black}{\bf(1.25)} & 2.30 \textcolor{black}{\bf(1.35)}& 1.56 \textcolor{black}{\bf(1.28)}\\   \hline                 
	\enddata
\end{deluxetable}


\section{LISM magnetic turbulence}\label{sec:LISM}
\begin{figure*}\centering
\includegraphics[width=0.8\textwidth]{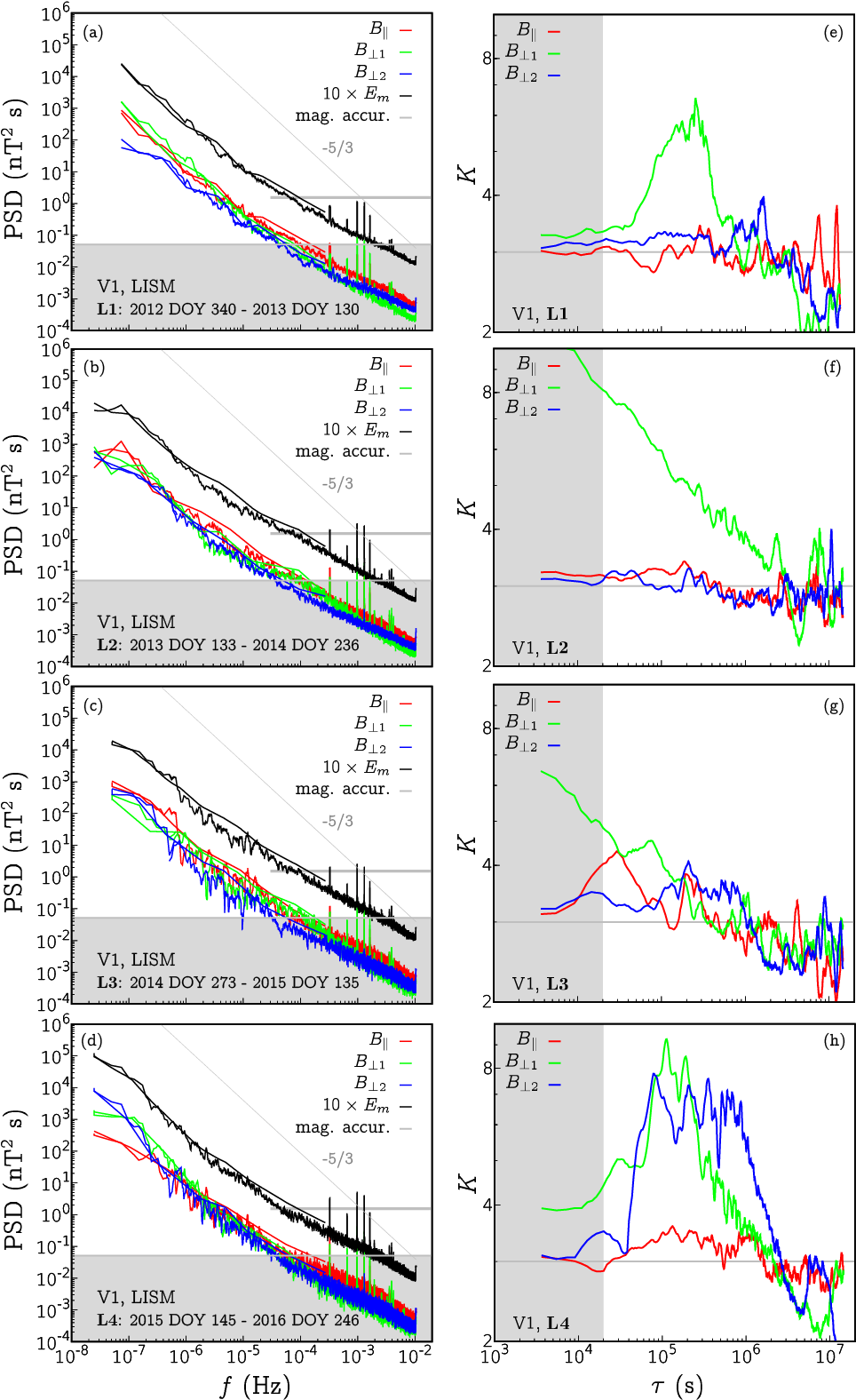}
 	\caption{Local Interstellar Medium. Left panels (a-d): power spectral density of magnetic field fluctuations. Right panels (e-h):  kurtosis of magnetic field increments for each field component.\label{fig:V1_LISM_ene}}
 \end{figure*}

\begin{figure}[h!]
 	\plotone{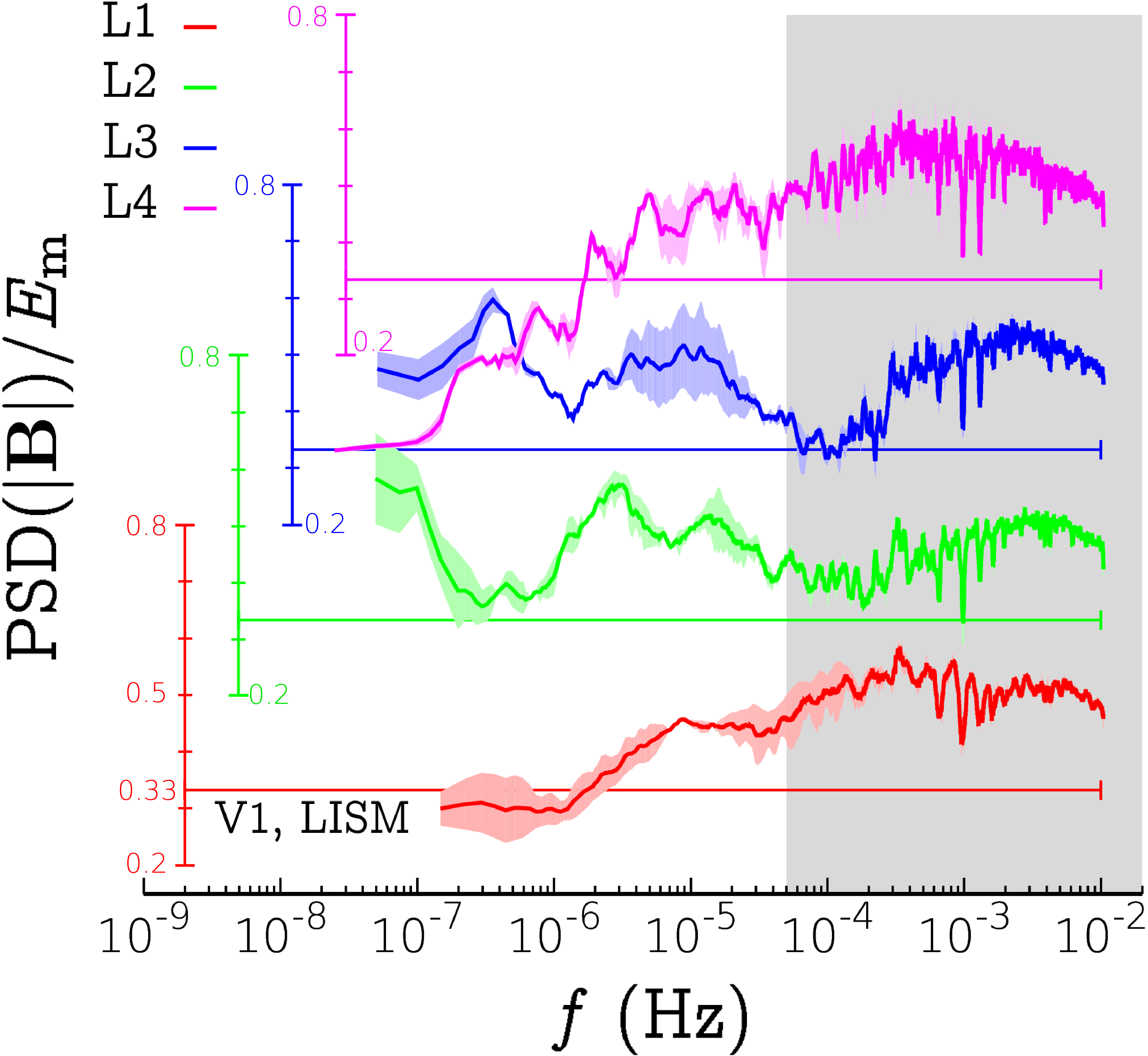}
 	\caption{Spectral compressibility  in the LISM. Average values computed through Eq. \ref{eq:compressibility} are shown in Table \ref{tab:LISM_fluc_statistics2_V1}. \label{fig:LISM_compressibility}}
 \end{figure}

Figure \ref{fig:V1_LISM_ene} shows the power spectra of the interstellar magnetic field in four V1 intervals \textbf{L1}--\textbf{L4}. \edit1{Anisotropy is shown in Figure \ref{fig:compress_anis_LISM_V1} in Appendix \ref{sec:App_B}}. As shown in the summary Table \ref{tab:LISM_fluc_statistics2_V1}, the intensity of magnetic fluctuations with respect to the background field ($B_{0\ \mathrm{LISM}}\approx 0.5$ nT) is nearly one order of magnitude smaller than in the IHS. The fluctuating magnetic energy increases significantly in the later periods \textbf{L3} and \textbf{L4}, and \textbf{L1} is the most quiet interval. There is little variation of the rms of parallel fluctuations among the intervals, even though it is higher during the central periods \textbf{L2} and \textbf{L3}. In fact, the central periods are the most compressible  ($C\approx0.5$). We observe a progressive increase of  transverse fluctuations in the $\perp_2$ (radial) direction. In fact, during \textbf{L4} magnetic fluctuations are primarily transverse, especially at the largest scales, as highlighted in Figure \ref{fig:LISM_compressibility} (pink curve) and in Figure \ref{fig:compress_anis_LISM_V1}(h). This fact was first pointed out by \cite{burlaga2018}.

Power spectra shown in the left panels of Figure \ref{fig:V1_LISM_ene} contain five frequency decades, a range unexplored so far.  The figure shows the noise level corresponding to a white noise with 0.04 nT amplitude (gray bands), i. e. $P_{noise}\approx0.05\ \mathrm{nT}^2$s.  
\cite{burlaga2018} indicate that noise may affect the data at $f\gtrsim4\times 10^{-5}$ Hz, which is consistent with the spectral flattening we observe. \edit1{It is interesting to note, however, that the level of anisotropy in this range remains high, at values around 0.45, and that different profiles are shown across intervals (Figure \ref{fig:LISM_compressibility}), which might be indicative of a smaller noise than estimated}.

 However, one can observe a spectral flattening (or a small bump) occurring, for all periods, in the range $10^{-6}<f<10^{-5}$ Hz. Moreover, the ion cyclotron frequency is smaller than in the IHS, $f_{ci,\mathrm{LISM}}\approx 10^{-4}$ Hz. At lower frequencies, the energy decays as a power law with spectral index close to the Kolmogorov's -5/3 value. Values reported in Table \ref{tab:LISM_fluc_statistics2_V1} have been computed in the range $5\times 10^{-8}<f<3\times10^{-6}$ Hz. As usual, errors indicate the discrepancy between the three spectral estimation techniques and it is higher than in the IHS, as only  the first decade is considered. 
It should be noted that in the interval \textbf{L2} a  spectral flattening occurs at the lower  frequencies ($f<3\times10^{-7}$ Hz). This may indicate that the turbulence is young and locally generated or, alternatively, affected by local structures as shocks. During \textbf{L4}, magnetic fluctuations seem to change nature, since transverse fluctuations (both along $\perp_1$ and $\perp_2$) become dominant and a clear power-law decay of energy is observed, with spectral index about -1.9 which is in part due to some rapid shears in  the signal. The parallel cascade is much slower, $\alpha\approx-1.4$.

The LISM turbulence is expected to show  the features of intermittency, which has not been considered yet in the literature. Our results are shown in Figure \ref{fig:V1_LISM_ene} (e-h), which show the scale-dependent kurtosis of magnetic increments as was done for the IHS data in the previous section.  We see that there is no intermittency at time scales $\tau\gtrsim 10^{6}$ s, for all intervals, since the kurtosis is smaller than three. At smaller scales, intermittency is observed for the $\perp_1$ component only, in the first three periods. During \textbf{L4} instead, a significant increase of the kurtosis is observed for both $\perp_1$ and $\perp_2$, in the range $10^5\lesssim\tau\lesssim3\times10^6$ s, where the energy cascade is fast. It should be noticed that in the noisy range ($\tau\lesssim10^4$ s) the statistic returns to Gaussian. Moreover, both during \textbf{L1} and \textbf{L4}, reduction of $K$ starts at larger scales, at about $\tau\approx 3\times10^5$ s.  This corresponds to the flattening observed in the power spectrum. Though it is possible that such reduction is an artifact due to data uncertainty,  it does not occur systematically, so at the current state of our analysis we do not exclude physical reasons. Increments of parallel fluctuations never show strong intermittency (differently to what was observed in the IHS).
 The structure function exponents reported in Table \ref{tab:LISM_fluc_statistics2_V1} have been computed in the range $\tau=[5\times10^5,5\times10^6]$ s. The absence of intermittency at larger scales may be a result of the passage of shocks - which could have made the  pristine interstellar fluctuations more Gaussian - or a signature of locally produced turbulence. The possibility that MHD waves could be transmitted from the IHS to the LISM has also been hypothesized \citep{zank2017}.



\begin{deluxetable}{l|cccc}[]
	\tablecaption{Magnetic field fluctuations properties in the LISM. \label{tab:LISM_fluc_statistics2_V1}}
	\tablecolumns{5}
	\tablewidth{0pt}
	\tablehead{
		\colhead{} & 	\textbf{L1} & \textbf{L2} & \textbf{L3} & \textbf{L4}	}
	\startdata 
	$  E_\mathrm{m} \ \mathrm{(nT^2)}$           & 2.84$\times10^{-4}$   &  2.38$\times10^{-4}$   &  3.72$\times10^{-4}$  & 6.86$\times10^{-4}$   \\	
	$C_\mathrm{1}$                 & 0.30& 0.40& 0.48 & 0.12 \\
	$C_\mathrm{2}$                 & 0.41& 0.54&0.65& 0.12\\
	$I_\parallel$                 & 0.014&0.018 & 0.023& 0.015 \\
	$I_{\perp1}$                 & 0.019& 0.016& 0.014& 0.030 \\
	$I_{\perp2}$                  & 0.009& 0.012& 0.014&  0.038\\
	$I$                            & 0.029&0.030& 0.035&  0.057\\ 
	$\delta B_\mathrm{mv}$         & 0.013& 0.011&0.015& 0.019 \\ 
	$\theta_{mv}$  & 65$^\circ$&23$^\circ$& 173$^\circ$&98$^\circ$\\  \hline
	$\alpha$               & -1.57$\pm$0.05& -1.65$\pm$0.10& -1.55$\pm$0.10&-1.77$\pm$0.10   \\
	$\alpha_\mathrm{\parallel}$       &-1.59$\pm$0.05  & -1.60$\pm$0.10& -1.57$\pm$0.10&-1.40$\pm$0.06 \\ 
	$\alpha_\mathrm{\perp}$           &-1.54$\pm$0.02& -1.60$\pm$0.05& -1.54$\pm$0.10&-1.90$\pm$0.05 	\\     \hline 
	$ \zeta_\mathrm{1, \parallel}$    \textcolor{black}{$ (\zeta_\mathrm{1, \parallel}^{\mathrm{ess}})$}               & 0.18  \textcolor{black}{\bf(0.33)} & 0.15  \textcolor{black}{\bf(0.34)}&  0.14  \textcolor{black}{\bf(0.38)} & 0.15  \textcolor{black}{\bf(0.35)} \\  
	$ \zeta_\mathrm{2, \parallel}$   \textcolor{black}{$ (\zeta_\mathrm{2, \parallel}^{\mathrm{ess}})$}                 & 0.36  \textcolor{black}{\bf(0.67)} & 0.28  \textcolor{black}{\bf(0.68)}& 0.27  \textcolor{black}{\bf(0.71)}  & 0.29  \textcolor{black}{\bf(0.68)}\\
	$ \zeta_\mathrm{3, \parallel}$   \textcolor{black}{$ (\zeta_\mathrm{3, \parallel}^{\mathrm{ess}})$}                 & 0.55 (1) & 0.40 (1)& 0.39  (1) & 0.43 (1)\\  
	$ \zeta_\mathrm{4, \parallel}$    \textcolor{black}{$ (\zeta_\mathrm{4, \parallel}^{\mathrm{ess}})$}                & 0.73  \textcolor{black}{\bf(1.32)} &   0.51  \textcolor{black}{\bf(1.31)}& 0.49  \textcolor{black}{\bf(1.25)} & 0.56  \textcolor{black}{\bf(1.30)}\\       \hline  
	$ \zeta_\mathrm{1, \perp}$    \textcolor{black}{$ (\zeta_\mathrm{1, \perp}^{\mathrm{ess}})$}               & 0.35  \textcolor{black}{\bf(0.36)} & 0.25  \textcolor{black}{\bf(0.38)}& 0.21  \textcolor{black}{\bf(0.40)}   & 0.40  \textcolor{black}{\bf(0.43)}\\              
	$ \zeta_\mathrm{2, \perp}$    \textcolor{black}{$ (\zeta_\mathrm{2, \perp}^{\mathrm{ess}})$}                & 0.67  \textcolor{black}{\bf(0.71)} & 0.48  \textcolor{black}{\bf(0.71)}& 0.37   \textcolor{black}{\bf(0.72)} & 0.72  \textcolor{black}{\bf(0.76)}\\       
	$ \zeta_\mathrm{3, \perp}$   \textcolor{black}{$ (\zeta_\mathrm{3, \perp}^{\mathrm{ess}})$}                   & 0.96 (1)& 0.69 (1)&  0.5 (1)& 0.96  (1)\\          
	$ \zeta_\mathrm{4, \perp}$   \textcolor{black}{$ (\zeta_\mathrm{4, \perp}^{\mathrm{ess}})$}                   & 1.21  \textcolor{black}{\bf(1.23)} & 0.88  \textcolor{black}{\bf(1.24)}& 0.59   \textcolor{black}{\bf(1.24)} & 1.12  \textcolor{black}{\bf(1.17)}\\   \hline               
	\enddata
\end{deluxetable}

\section{Summary and final remarks}\label{sec:conclusions}
This work provides a broadband spectral and high-order statistical analyses of magnetic field fluctuations in the IHS and LISM which give new insights into the properties of turbulence and its evolution. Our aim was to investigate the existence of different regimes of turbulent fluctuations, characterize them and describe their evolution in time and space. 

We considered 12 data sets at different times and latitudes, after 2009, between 88 AU and 135 AU. Analysis of high-resolution (48 s) \textit{in situ} measurements from both \textit{Voyager 1} and \textit{2} spacecraft with the proposed advanced spectral-estimation techniques made it possible to investigate the evolution of fluctuations across more than five frequency decades ($10^{-8}<f<10^{-2}$ Hz), a range of scales which has not been explored in the literature so far. We focused our attention on the energy cascade, the compressible nature, anisotropy and intermittency of magnetic fluctuations.

In the IHS at V2, we  identified the \textit{Energy-Injection} and \textit{Inertial-Cascade} regimes. Instead, the signatures of the kinetic regime were expected to show up in the last decade of the frequency spectrum ($10^{-3}<f<10^{-2}$ Hz), but at present \edit1{the unknown exact level of noise in data does not allow us to discuss this regime. However, we believe that physical phenomena may still be detected in this range so that it deserves further study}. 

The EI range  is featured by a 1/f power law decay of magnetic energy, non-intermittent statistics of magnetic increments, and low compressibility. Its frequency extent depends on the observational interval considered, and seems to be wider in the unipolar periods. The first relevant scale highlighted in this study corresponds the EI/IC spectral break, and could not be just related to the  nominal spacing of sectors ($\approx$2 AU), nor to the age of fluctuations generated or affected by the termination shock. In the unipolar periods, for instance, the spectral break occurs at  $f\approx10^{-5}$ Hz, corresponding to a spatial scale of about 0.15 AU along the wind direction. The originally more \edit1{Alfv{\`e}nic nature of unipolar regions may explain the observed difference between SHS and UHS periods}.

The IC regime is characterized by (i) steepening of  power spectra towards values of spectral index close to the Kolmogorov's value of -1.67, (ii) rapid growth of the kurtosis of magnetic field increments, and  (iii) a clearly-defined power law decay of the third-order structure function $S_3$, with  typical exponents of MHD turbulence. It can be concluded that unipolar periods showed a faster energy decay than sectored periods. The second typical scale in the V2 analysis was observed at $f\approx10^{-4}$ Hz ($\ell\approx5\times10^{-3}$ AU). Here, a  spectral knee occurs mainly for $\delta B_\parallel$, and the maximum anisotropy and compressibility is observed.  

Fluctuations in the broad unipolar regions traveled by V1 before 2011.5 show a marked dominance of $\delta B_\parallel$. Parallel energy experiences a spectral break separating the large-scale regime with slow 1/f decay from the fast  regime with $\alpha\approx-2.5$. Interpretation of V1 spectra is challenging due to the lack of accurate plasma velocity data and slow wind conditions, which do not allow us to compute the wavenumbers. It is possible that the parallel-wavenumber components due to Alfv{\'e}nic or fast magnetosonic fluctuations dominate the spacecraft-frame frequency, but at present we cannot verify this hypothesis. We showed that  V1 fluctuations are intermittent with a power-law increase of kurtosis anticipating a spectral break. A similar trend is also observed for the spectral compressibility, which reaches its maximum at the break frequency.

Finally, we have analyzed four LISM intervals. Even in this case, we used  48 s  data to extend the range of frequencies considered in past literature studies and improve the accuracy of spectral estimates. We note that the level of compressible fluctuations is not higher than 0.6, and confirm the recently observed change of nature of turbulence during 2015/2016. Such transition consists mainly of an  increase in perpendicular energy (especially, in the $\perp_2$ component). Moreover, in all intervals we observed a spectral flattening resembling a small bump,  for $10^{-6}\lesssim f\lesssim10^{-5}$ Hz. This bump also corresponds to an increase in compressibility. Its nature should be further investigated. Intermittency is mainly observed in transverse fluctuations. Such rapid increase begins at a spacecraft frequency of $3\times10^{-7}$ Hz. \textcolor{black}{It is then plausible that the observed LISM turbulence is locally modified by the periodic passage of shocks.}

We expect our results to provide additional constraints on numerical and theoretical models of the outer heliosphere and, hopefully, shed light onto transport properties of energetic particles in these regions of space.

\FloatBarrier

\acknowledgments
\edit1{We thank the 
anonymous referee for helpful comments and suggestions.} The authors acknowledge support from the MISTI-Seeds MITOR project ``Spectral analysis of the solar wind beyond the termination shock - interstellar medium/heliosphere interactions'' 2015-2016.    FF was also supported by the post-doctoral grant ``FOIFLUT'' 37/17/F/AR-B. NP was supported, in part, by NASA grants NNX14AJ53G, NNX16AG83G, and 80NSSC18K1649NS, and NSF PRAC award OAC-1811176. JDR was supported under NASA contract 959203 from JPL to MIT.  Computational resources for spectral and statistical analysis were provided by HPC@POLITO (\url{http://www.hpc.polito.it}).  

\appendix
\section{Methods for spectral analysis}\label{sec:App_A}

\begin{figure*}
    \centering
    \includegraphics[width=\textwidth]{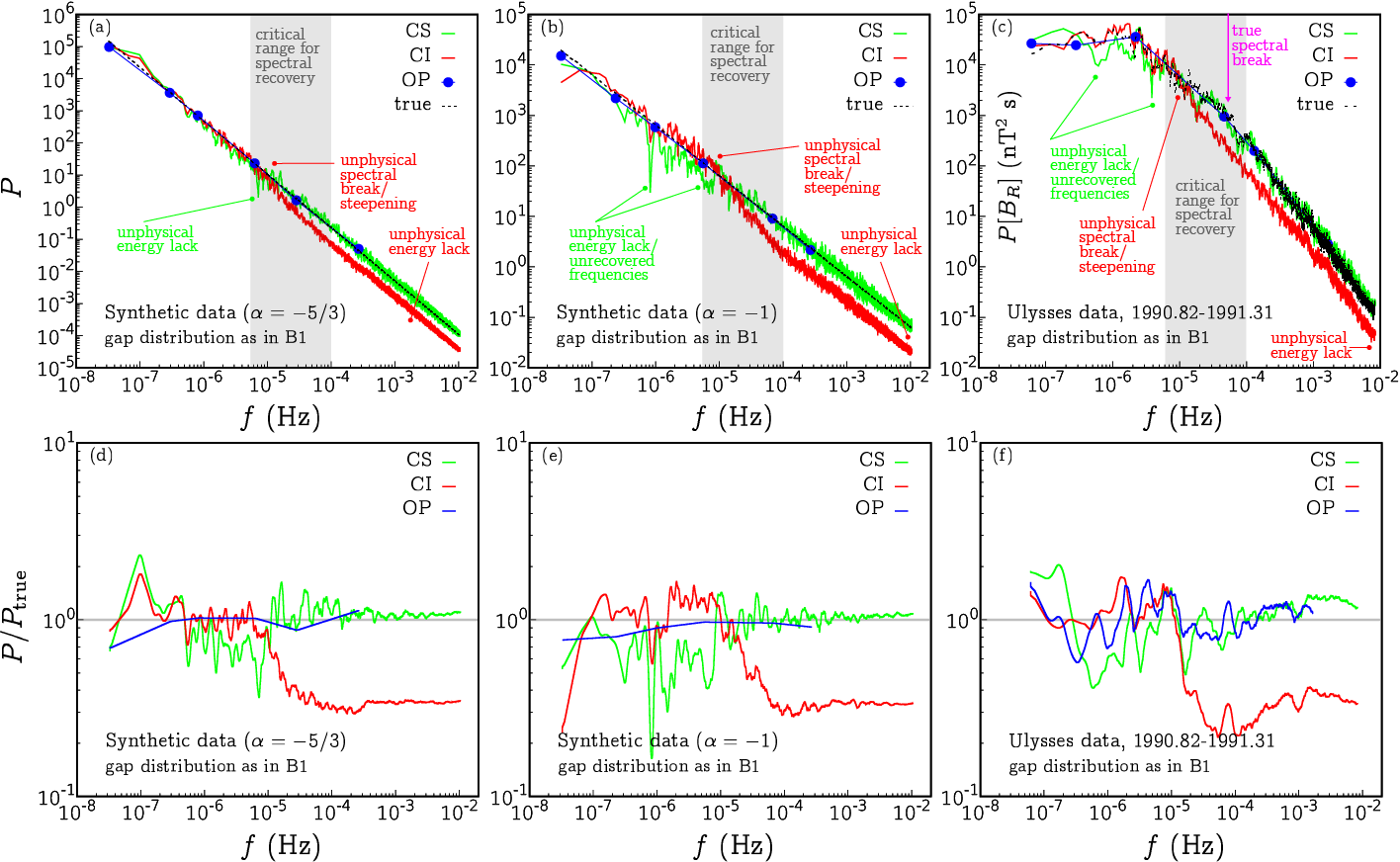}
    \caption{Testing of spectral estimation techniques on  synthetic turbulence data with constant spectral index, $\alpha=-5/3$ (a,d); synthetic turbulence data with  $\alpha=-1$ (b,e); \textit{Ulysses} data (1990.82-1991.31, gap-free) (c,f). Top panels show the PSD, bottom panels show the ratio between the estimated spectrum and the true one. In these tests, data points have been artificially removed with the same gap distribution of the period \textbf{B1} (68\% of missing points; longest gap: 132 h; longest gap-free subset length: 16.5 h; average gap-free subset length: 3.25 hours). \label{fig:methods_test}}
\end{figure*}

Computing power spectra over a broad range of frequencies in the outer heliosphere is challenging because of the sparsity of the 48 s data (70\% of magnetic field are missing).  \citet{fraternale2016},  \citet{gallana2016}, \citet{iovieno2016} and \citet{fraternale2017phd} have demonstrated that a careful application of different, independent, techniques makes it possible to recover the spectrum with proper accuracy  (e.g., with the accuracy of 10\% or lower in the spectral index. \edit1{The techniques description and numerical codes have been provided in the above references. Here, we briefly recall them with focus on their specific application to Voyager data in the IHS, together with three examples of tests conducted on contiguous data sets artificially gapped (synthetic turbulence and Ulysses data, see Figure \ref{fig:methods_test}).}  

\textit{1. Correlation method with linear data interpolation}  (CI): the PSD is obtained as the Fourier transform of the two-point auto-covariance function computed from linearly-interpolated data. It recovers the spectrum  well in the low-frequency range, i. e. at frequencies smaller than the typical frequency of large data gaps ($f_{gap}\approx 2\times10^{-5}$ Hz). Due to the low-pass effect of the linear interpolation, a spectral leakage  is observed at higher frequencies and  spectral exponents are typically overestimated \edit1{(70\% energy loss, see bottom panels in Figure \ref{fig:methods_test}, red curves)}.

\textit{2. Compressed sensing spectral estimation} (CS): a recent paradigm  we adopted from the signal processing and telecommunication area \citep{candes2006a, candes2006b,donoho2006}. This method does not interpolate data. It allows to recover exactly sparse
signals (signals with few nonzero frequencies) even if fewer data points than those required by the Shannon's principle are available.
Testing CS on turbulent, gapped, data sets, we found it recovers well the spectrum, especially the high-frequency range ($f\gtrsim f_\mathrm{gap}$). Depending on the interval considered, it may lack accuracy in the neighborhood of $f_{gap}$. This typically shows up as a lack of energy around $f\in[3\times 10^{-6},10^{-5}]$ Hz (green curves in Figure \ref{fig:methods_test}). Also, a small peak around  $f_\mathrm{gap}$ is sometimes observed (see the case of \textbf{SHS2}, Figure \ref{fig:V2_IHS_ene}d). CS has also been recently exploited for the analysis of the spectrum of magnetospheric intervals out of a Kelvin-Helmholtz-instability event observed by the Magnetospheric Multiscale (MMS) mission in the Earth magnetosphere \citep{sorriso2019}.

\textit{3. Optimization method} (OP): it is a simple algorithm which aims at minimizing errors in the CI analysis. This method is based on a genetic algorithm \citep[our code is based on the open-source code PIKAIA]{charbonneau1995} which returns the piecewise-linear model spectrum, $P_\mathrm{OP}(f)$ as a result. Starting from an initial model spectrum, a synthetic signal is obtained by inverse Fourier transform. From this synthetic set, some data are removed according to the gap distribution in the \emph{Voyager} data set.  The power spectrum of this gapped synthetic set  is successively computed by using the CI technique ($P_\mathrm{sy,~CI}$). The result is then compared to the CI spectrum of the Voyager data set ($P_\mathrm{voy,~CI}$). The difference between the two spectra ($F=\sum_i |P_\mathrm{sy,~CI}(f_i)-P_\mathrm{voy,~CI}(f_i)|$) is minimized by the optimization algorithm which, at each generation, modifies the control points in the model spectrum which is an approximation of the true spectrum of the \textit{Voyager} signal \citep[][Chapt. 4]{fraternale2017phd}. This method helps to estimate the error of the linear interpolation. Heuristically, OP proved to work well for statistically-homogeneous data sets representing physical phenomena with a continuous spectrum distributed over a broad range of scales. It cannot represent peaks in the spectrum but, if any, they are well identified by the other methods above (Figure \ref{fig:methods_test}, blue curves).

\edit1{\textit{4. Gap-free subsets}  (SS): to check results in the high-frequency range, we compute the averaged spectrum of contiguous subsets. Given the gap distribution of IHS and LISM \textit{Voyager} data, it allows to see the frequency range $10^{-4}\lesssim f \lesssim 10^{-2}$ Hz. In fact, for all intervals considered in this study, the average subset length is $3.7\pm0.6$ hours in the IHS and $2.49\pm0.07$ in the LISM; the maximal length is $14.4\pm3.3$ h in the IHS and $10.3\pm2.3$ in the LISM. On average, ensambles include $520$ subsets (Figures \ref{fig:compress_anis_IHS_V2}, \ref{fig:compress_anis_IHS_V1} and \ref{fig:compress_anis_LISM_V1}). }

Power spectra displayed in Figures \ref{fig:V2_IHS_ene}, \ref{fig:V1_IHS_ene}, and \ref{fig:V1_LISM_ene} are build by using the average result of CI and CS  for $f<10^{-5}$ Hz, and for $f>10^{-5}$ Hz  \edit1{CS} only. The results of the OP method is also shown in all PSD pictures by a continuous smooth curve. In general, a very good agreement is observed. Note also that the CI and CS spectra are smoothed by means of a moving average with a constant-width window in the logarithmic space (i.e. the averaging points increase linearly with the frequency). This smoothing has no effect on the level of power and on the spectral index.

Figure \ref{fig:methods_test}  shows the application of methods CI (red curves), CS (green curves), OP (blue curves) to three data sets, where data have been artificially removed according to the same gap distribution of the period \textbf{B1}. The first data set is a synthetic turbulence set (obtained by inverse transform of a given power spectrum - black line - using random phases of the Fourier coefficients) with  a constant index  $\alpha=-5/3$ (Figure \ref{fig:methods_test}a). In the second case $\alpha=-1$, which represents a harder test case (panel b). In the third case, we used gap-free \textit{Ulysses} data in the period 1990.82-1991.31. Panels (d)-(f) show the (smoothed) ratio  between the estimated and the true spectrum $P/P_\mathrm{true}$. 

It should be noticed that - due to the gap distribution of \textit{Voyager} magnetic field time series  in the IHS and LISM - the frequency range where spectral estimation is more critical is around $10^{-5}<f<10^{-4}$ Hz. The lower bound corresponds to $f_\mathrm{gap}\approx2\times10^{-2}$ Hz, the higher is linked to the length of contiguous subsets. 

\FloatBarrier

\section{Variance anisotropy}\label{sec:App_B}

\begin{figure}
    \centering
    \includegraphics[width=0.73\textwidth]{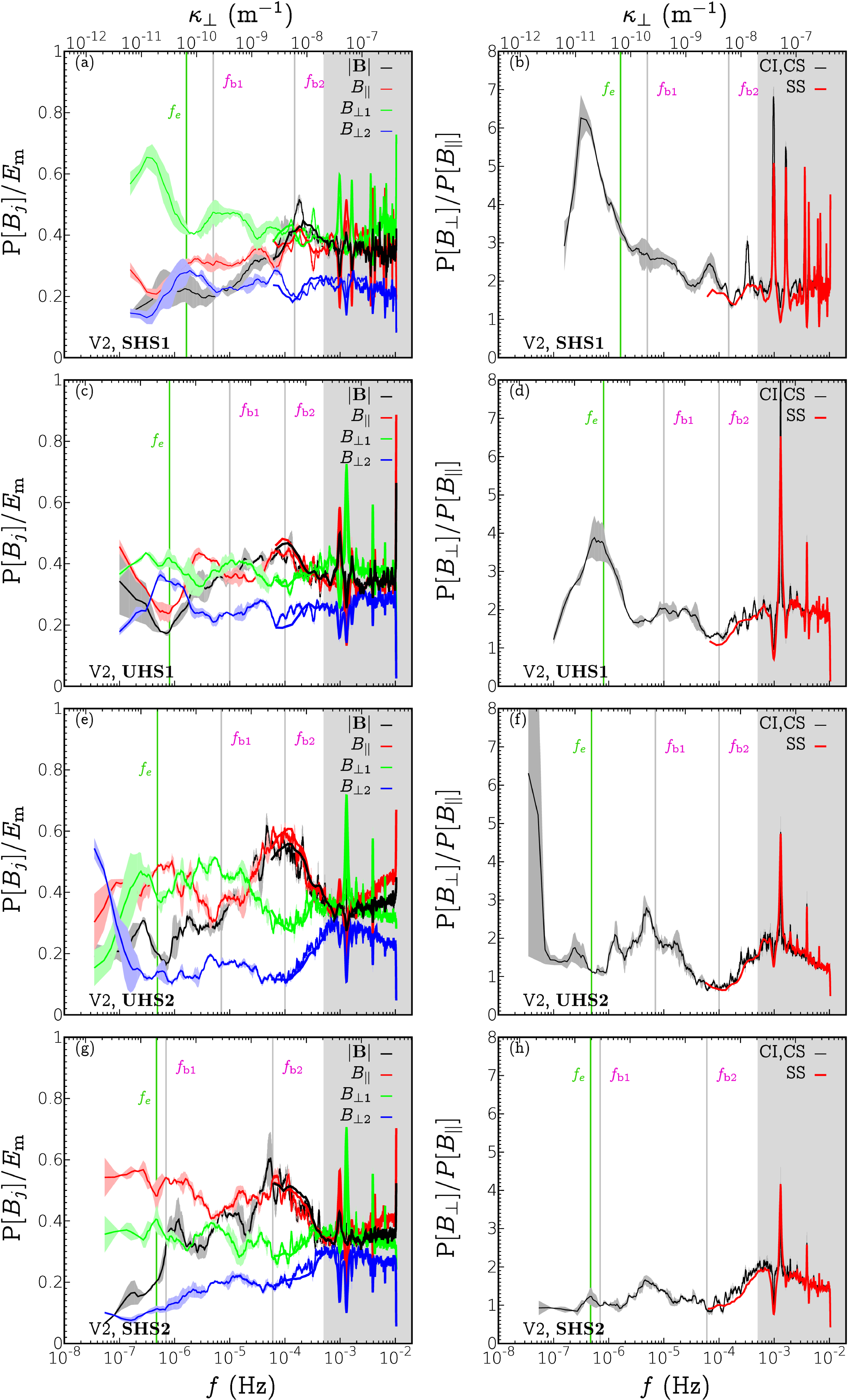}\vskip-5pt
    \caption{\edit1{Spectral variance anisotropy and compressibility in the IHS at \textit{Voyager 2}. Top to bottom: \textbf{SHS1}, \textbf{UHS1}, \textbf{UHS2} and \textbf{SHS2} intervals. Left panels: anisotropy computed as $P[B_j]/E_m$, where  $B_j=\{B_\parallel,B_{\perp1},B_{\perp2}, |\mathbf{B}|\}$ and $E_m$ is the trace. The black curve represents the spectral compressibility proxy, based on the magnetic field magnitude. The thick continuous lines stand for the SS method (gap-free sub-sets), the thin curves show the average result of methods CI and CS, together with error bands. Right panels: ratio between B-perpendicular and B-parallel energy. Here,  black lines show the average result of methods CI and CS, (with error bands) and red lines show the result from contiguous sub-sets (SS). The peaks in the last decade are due to instrumental interference.}    \label{fig:compress_anis_IHS_V2}}
\end{figure}

\begin{figure}
    \centering
    \includegraphics[width=0.75\textwidth]{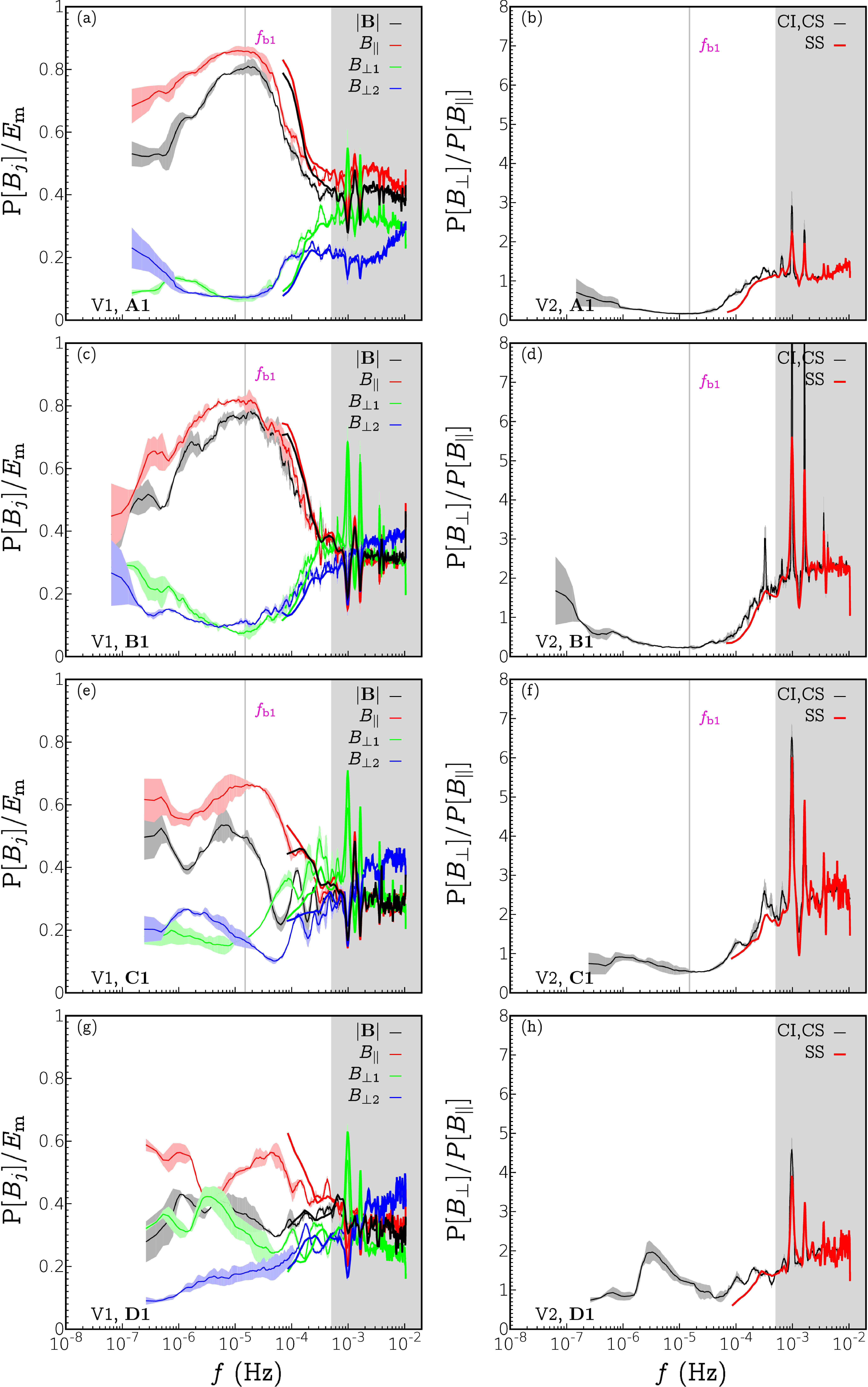}
    \caption{\edit1{Spectral variance anisotropy and compressibility in the IHS at \textit{Voyager 1}. Top to bottom: \textbf{A1}, \textbf{B1}, \textbf{C1} and \textbf{D1} intervals. Panels description as in caption of Figure \ref{fig:compress_anis_IHS_V2}.}\label{fig:compress_anis_IHS_V1}}
    
\end{figure}

\begin{figure*}
    \centering
    \includegraphics[width=0.75\textwidth]{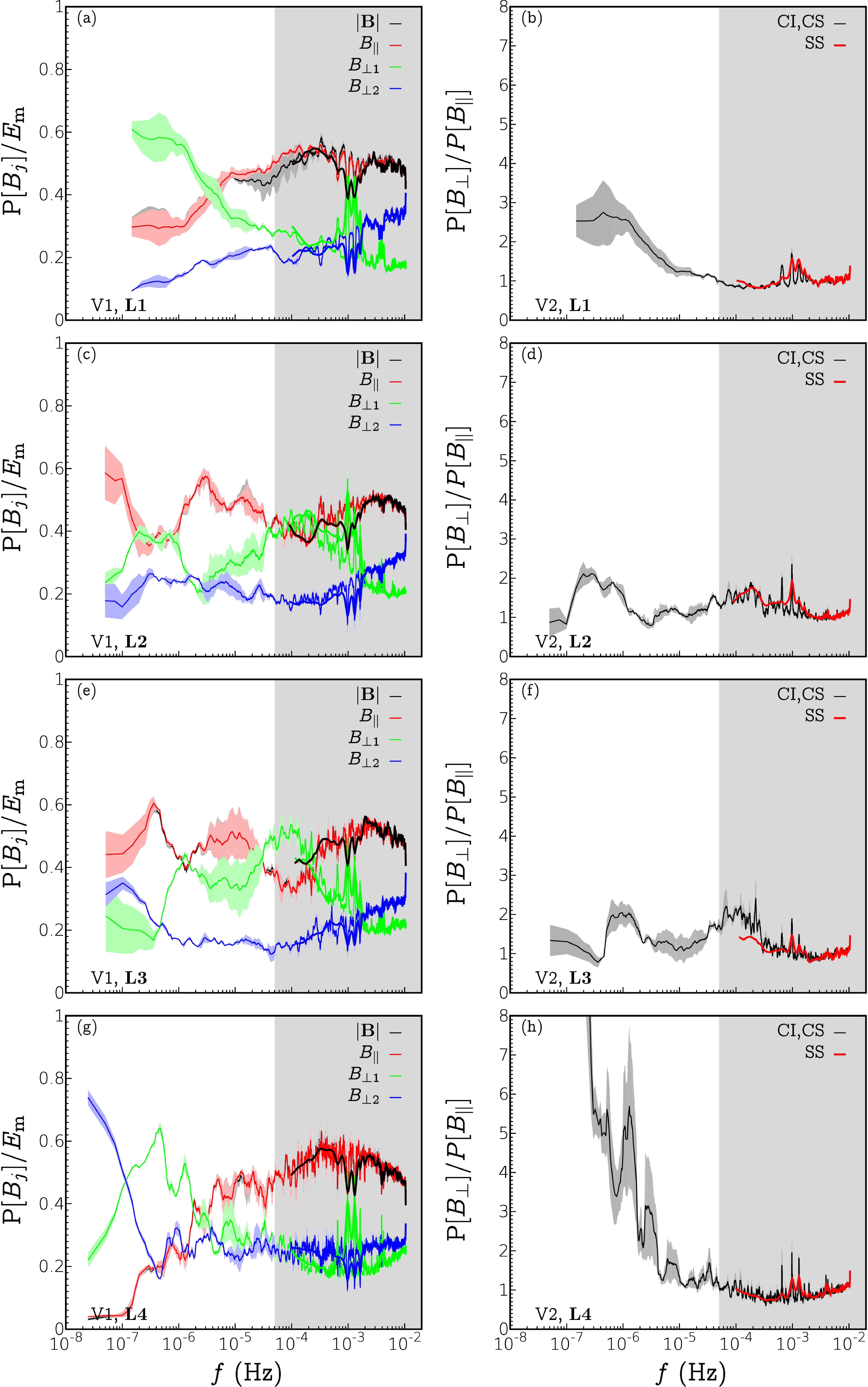}
    \caption{\edit1{Spectral variance anisotropy and compressibility in the LISM at \textit{Voyager 1}. Top to bottom: \textbf{L1}, \textbf{L2}, \textbf{L3} and \textbf{L4} intervals. Panels description as in caption of Figure \ref{fig:compress_anis_IHS_V2}.}\label{fig:compress_anis_LISM_V1}}
\end{figure*}

\edit1{Figures \ref{fig:compress_anis_IHS_V2},\ref{fig:compress_anis_IHS_V1},\ref{fig:compress_anis_LISM_V1} show the frequency-space anisotropy for all IHS and LISM intervals presented in this study. Left panels show the fractional energy of each magnetic field component with respect to the trace, that is $P[B_j]/E_m$. The $|\mathbf{B}|$ case, black curves, is the spectral compressibility proxy already shown in Figures \ref{fig:spec_compressibility} and \ref{fig:LISM_compressibility}. 
From the left panels, one can see that the $\delta B_\parallel$ curve (red) follows with the $|\textbf{B}|$ curve (black) in most cases, especially for $f>10^{-5}$ Hz (a worse agreement is observed for \textbf{C1} and \textbf{D1} periods, which are shorter than others). Some discrepancy between the two curves is observed in sectored regions in the low-frequency regime: this is due to tangential discontinuities in correspondence of sector boundary crossings (here, $B_\parallel$ changes sign). Such reversals are seen in the spectrum as large-amplitude $\delta B_\parallel$ fluctuations, but are not related to compressions. In fact, they are not accounted for by the $P[|\mathbf{B}|]/E_m$ indicator.
The peak of compressibility and anisotropy occurs at $f_{b2}$ for V2 and at $f_{b1}$ for V1. Note also that a significant level of anisotropy is retained in the gray region, where data may be affected by noise, especially in the LISM (Figure \ref{fig:compress_anis_LISM_V1}). 
}

\FloatBarrier

\FloatBarrier


\end{document}